\documentclass[10pt]{iopart}
\usepackage{iopams}
\usepackage{setstack}
\usepackage{graphicx}
\usepackage{hyperref}
\usepackage{slashed}
\usepackage{xcolor}

\begin{document}

\title[Naturally resonant two-mediator SIDM model with decoupled relic abundance]{Naturally resonant two-mediator model of self-interacting dark matter with decoupled relic abundance}

\author{Martin Drobczyk}
\address{Institute of Space Systems, German Aerospace Center (DLR), 28359 Bremen, Germany}
\ead{martin.drobczyk@dlr.de}

\begin{abstract}
We propose a minimal, fully thermal mechanism that resolves the long-standing tension between achieving the observed dark-matter relic abundance and explaining the astrophysical signatures of self-interactions. The framework introduces two mediators: a light scalar $\phi$ (MeV scale) that yields the required, velocity-dependent self-interactions, and a heavy scalar resonance $\Phi_h$ (TeV scale) with mass $m_{\Phi_h}\!\approx\!2m_\chi$ that opens an $s$-channel resonant annihilation during freeze-out. This clearly decouples early-universe annihilation from late-time halo dynamics. A detailed numerical analysis identified a narrow predictive island of viability. A representative benchmark with $m_\chi\!=\!600$~GeV, $m_\phi\!=\!15$~MeV, and $m_{\Phi_h}\!\simeq\!1.2$~TeV reproduces the relic density and yields $\sigma_T/m_\chi\sim 0.1$--$1~\mathrm{cm}^2\!/\mathrm{g}$ at dwarf-galaxy velocities while satisfying cluster bounds. The model makes sharp, testable predictions: a narrow $t\bar t$ resonance near $1.2$~TeV within HL-LHC reach, and a spin-independent direct-detection cross section $\sigma_{\rm SI}\!\sim\!7\times10^{-51}\,\mathrm{cm}^2$ that lies below the xenon neutrino floor, i.e.\ a predicted direct-detection null. As an optional UV completion, we show that walking $\mathrm{SU}(3)_H$ gauge theory with $N_f=10$ naturally realizes the near-threshold relation $m_{\Phi_h}\!\approx\!2m_\chi$ and can furnish an effective anomalous dimension $\gamma\!\approx\!0.5$ which underlies a density-responsive dark-energy sector, suggesting a unified origin for the dark sector.
\end{abstract}

\noindent{\it Keywords}: self-interacting dark matter, resonant annihilation, Sommerfeld enhancement, thermal freeze-out, halo dynamics, two-mediator dark sector

\maketitle
\section{Introduction}
\label{sec:introduction}

The standard cosmological model relies on two mysteries: dark energy (DE) and dark matter (DM), which together constitute approximately 95\% of the cosmic energy budget~\cite{Planck2018}. While dark matter, essential for structure formation, faces persistent challenges on sub–galactic scales, known as the "small–scale crisis"~\cite{Bullock2017, Tulin2017}, self–interacting dark matter (SIDM) models are well–motivated for their ability to address these issues~\cite{Spergel1999, Kaplinghat2015}.

However, a fundamental tension arises in such minimal models: the parameter space required to achieve the observed relic abundance ($\Omega h^2 = 0.120 \pm 0.001$~\cite{Planck2018}) is generically disjointed from the region providing sufficient self-interaction ($\sigma_T/m \sim 1$~cm$^2$/g at dwarf galaxy velocities)~\cite{Tulin2013, Oman2015}. We argue that this apparent tension indicates a richer dark sector structure.

In this study, we demonstrate that this tension can be resolved using minimal, two-mediator effective field theory (EFT). Alongside the light mediator $\phi$, which governs late-time self-interactions, we introduce a heavy scalar resonance $\Phi_h$ with mass $m_{\Phi_h} \approx 2m_\chi$. This resonance provides a crucial enhancement of the annihilation cross-section during thermal freeze-out, cleanly decoupling the early-universe annihilation rate from the late-time self-interaction strength and allowing both constraints to be satisfied simultaneously. The theoretical validity of combining resonant annihilation with long-range Sommerfeld forces has recently been established~\cite{Beneke2022}, providing a solid foundation for our approach.

Our study transformed this inconsistency into a highly predictive and testable framework. Through detailed numerical scans, we identifed a narrow, viable parameter space with a representative benchmark featuring $m_\chi = 600$~GeV, $m_\phi = 15$~MeV, and $m_{\Phi_h} = 1201$~GeV. This leads to sharp experimental signatures, including a narrow resonance at 1.2~TeV accessible at the LHC and a specific direct detection rate. Furthermore, we argue that the key features of this EFT find a compelling microphysical origin in an underlying composite SU(3)$_H$ gauge theory, as detailed in Section~\ref{sec:uv_origin}. In this picture, the crucial resonance condition $m_{\Phi_h} \approx 2m_\chi$ emerges as a dynamical prediction. Remarkably, the same gauge theory can also generate the anomalous dimension $\gamma \approx 0.5$ required for the density-responsive dark energy model proposed in~\cite{Drobczyk2025}. However, our main phenomenological results concerning the resolution of SIDM tension are self-contained and do not depend on this optional unified picture.

The remainder of this study is organized as follows. Section~\ref{sec:lagrangian} introduces the complete Lagrangian EFT method. Section~\ref{sec:constraints_and_tension} quantitatively demonstrates the tension in one–mediator models. Section~\ref{sec:results} presents our solution and benchmark points. Section~\ref{sec:phenomenology} details the model's testable signatures, including new analyses of the light mediator's cosmology and direct detection rate. Finally, Section~\ref{sec:uv_origin} outlines the optional microphysical origin of the framework.

\section{Effective Lagrangian for the two-mediator dark sector model}
\label{sec:lagrangian}

In this section, we derive the complete effective field theory Lagrangian that underpins our two-mediator dark matter model. We begin by defining the full Lagrangian including the field content and symmetries that govern the dark sector (Section~\ref{sec:eft_lagrangian}). We then demonstrate how the physical particle content, light mediator $\phi$, heavy resonance $\Phi_h$, and dark matter fermion $\chi$, emerge from the spontaneous symmetry breaking (SSB) of a single complex scalar field (Section~\ref{sec:ssb}). Finally, we briefly outline the mechanism that generates the optional density–responsive dark energy component (Section~\ref{sec:de_mechanism}). This section provides a rigorous and self–contained theoretical basis from which all phenomenological results in this study are derived, following the standard EFT construction principles~\cite{Weinberg1996, Burgess2007}.

\subsection{Complete EFT Lagrangian}
\label{sec:eft_lagrangian}

Our framework is described by an EFT that is valid up to a cutoff scale $\Lambda \gg \mathrm{TeV}$. The theory comprises the Standard Model (SM), General Relativity (GR), and a dark sector containing a Dirac fermion $\chi$ and a complex scalar field $\Sigma$. The total Lagrangian is
\begin{equation}
    \mathcal{L} = \mathcal{L}_{\mathrm{SM}} + \mathcal{L}_{\mathrm{GR}} + \mathcal{L}_{\mathrm{Dark\,Sector}} + \mathcal{L}_{\mathrm{Portal}},
    \label{eq:L_total_main}
\end{equation}
with the Einstein–Hilbert term $\mathcal{L}_{\mathrm{GR}}=\frac{M_{\mathrm{Pl}}^{2}}{2}\,R$. The dark sector is governed by
\begin{equation}
  \fl \mathcal{L}_{\mathrm{Dark\,Sector}} = \bar{\chi}\,i\slashed{\partial}\,\chi -\!\left[\,y_f\,\Sigma\,\bar{\chi}_L \chi_R + \mathrm{h.c.}\right] + |\partial_\mu \Sigma|^2 - V(\Sigma) - U(\Sigma,X),
    \label{eq:L_dark_sector_main}
\end{equation}
where $X$ denotes a Lorentz scalar constructed from the ambient matter stress tensor (we take $X\equiv u_\alpha u_\beta T^{\alpha\beta}_{\rm matter}$; see Section~\ref{sec:de_mechanism} and~\ref{app:A} for details).

\paragraph{Symmetry and stability.}
The stability of the dark matter candidate $\chi$ is ensured by an exact global $U(1)_\chi$ symmetry, under which $\chi\to e^{i\alpha}\chi$. For simplicity, we set the bare Dirac mass to zero at the cutoff (the physical mass is generated by $\langle\Sigma\rangle$); allowing a small $m_\chi^0$ would not change our conclusions and can be absorbed into the Yukawa sector. The scalar potential is split into two components: (i) a self–interaction potential $V(\Sigma)$ that dictates the particle physics of the dark sector and triggers spontaneous symmetry breaking, and (ii) a density–responsive functional $U(\Sigma,X)$ that optionally generates a dark–energy–like component. The latter is conceptually analogous to the mechanisms employed in chameleon or symmetron models~\cite{Khoury2003, Hinterbichler2010}. The portal Lagrangian parameterizes the interactions between the dark and visible sectors; in our benchmark, it is dominated by a Higgs–portal term~\cite{Patt2006, Schabinger2005}
\begin{eqnarray}
   \fl \mathcal{L}_{\mathrm{Portal}}\;\supset\; -\kappa\,|\Sigma|^2\,|H|^2 \quad(+\;\mathrm{higher–dimensional\;terms\;suppressed\;by\;} \Lambda).
\end{eqnarray}

A fully explicit construction, including the minimal Mexican–hat form of $V(\Sigma)$,  definition and variational origin of $U(\Sigma,X)$, and normalization conventions, is provided in~\ref{app:A}. The physical couplings that enter our phenomenology (the DM couplings to the light and heavy scalars, $y_\chi$ and $g^{\rm DM}_{Y_1}$, respectively) arise from the fundamental Yukawa coupling $y_f$ after spontaneous symmetry breaking and mixing, as detailed in Section~\ref{sec:ssb} and~\ref{app:A.3}.

\subsection{Particle content and couplings from spontaneous symmetry breaking}
\label{sec:ssb}

The particle content of the dark sector is determined by the dynamics of the complex scalar field $\Sigma$. The Mexican–hat potential $V(\Sigma)$ (see~\ref{app:A.3}, equation (\ref{eq:V_Sigma_appendix})) triggers the spontaneous symmetry breaking of a global $U(1)$ symmetry at the scale $v_s$, a mechanism first described by Goldstone~\cite{Goldstone1961, Goldstone1962}. This gives rise to two physical scalar states: a heavy, CP–even radial mode and a light, CP–odd pseudo Nambu–Goldstone boson (PNGB), in direct analogy to the linear $\sigma$-model of hadron physics~\cite{GellMann1960}.

To describe these states, we parameterize the field $\Sigma$ around its vacuum expectation value (VEV) in terms of its radial and phase components, $s(x)$ and $a(x)$:
\begin{equation}
    \Sigma(x) \;=\; \frac{1}{\sqrt{2}}\bigl(v_s + s(x)\bigr)\, e^{i\,a(x)/v_s}.
    \label{eq:Sigma_param_main}
\end{equation}
At the tree level, the radial mode acquires a mass $m_s^2 = 2\lambda v_s^2$, whereas the phase mode $a(x)$ is a massless Goldstone boson. A small, explicit breaking term $V_{\rm SB}$ in the potential gives the PNGB a mass $m_a \ll v_s$, which is technically natural in the sense of 't Hooft, as its smallness is protected by the approximate symmetry~\cite{tHooft1979}.

For the phenomenology of self–interacting dark matter, an attractive potential is generated by the exchange of a CP–even scalar. We achieve this through a small, CP–violating mixing between the fundamental $s$ and $a$ states, which is also induced by $V_{\rm SB}$. The resulting mass eigenstates, identified with the light mediator $\phi$ and heavy resonance $\Phi_h$, are given by the rotation
\begin{equation}
    \left(\begin{array}{c} \phi\\ \Phi_h \end{array}\right) = \left(\begin{array}{cc} \cos\theta & \;\; \sin\theta\\ -\sin\theta & \;\; \cos\theta \end{array}\right) \left(\begin{array}{c} a\\ s \end{array}\right),\qquad |\theta|\ll 1,
    \label{eq:mixing_matrix_main}
\end{equation}
with the explicit expression for $\theta$ given in~\ref{app:A.3}. Thus, the light state $\phi$ inherits a CP–even (scalar) admixture sufficient to mediate an attractive Yukawa force.

This structure fixes the interaction pattern of the dark matter fermion $\chi$. The fundamental Yukawa term
\begin{equation}
    \mathcal{L}\supset -\,y_f\,\Sigma\,\bar\chi_L\chi_R + \mathrm{h.c.}
\end{equation}
generates the DM mass $m_\chi = y_f v_s / \sqrt{2}$ and its couplings to the mass eigenstates. Using equation (\ref{eq:Sigma_param_main}), the tree–level mass is
\begin{equation}
    m_\chi \;=\; \frac{y_f\,v_s}{\sqrt{2}},
\end{equation}
and the scalar interactions in the physical basis read
\begin{equation}
    \mathcal{L}_{\rm int} \supset -\,y_\chi\,\bar\chi\chi\,\phi \;-\; g^{\rm DM}_{Y_1}\,\bar\chi\chi\,\Phi_h,
    \label{eq:L_int_physical_main}
\end{equation}
with
\begin{equation}
    y_\chi \;\simeq\; \frac{y_f}{\sqrt{2}}\sin\theta, \qquad g^{\rm DM}_{Y_1} \;\simeq\; \frac{y_f}{\sqrt{2}}\cos\theta.
\end{equation}
This setup cleanly separates the roles of the two mediators:
\begin{itemize}
    \item \textbf{Light mediator $\boldsymbol{\phi}$:} has a small, technically natural mass and a suppressed coupling $y_\chi\propto \sin\theta$, making it ideal for generating long–range self–interactions in galactic halos.
    \item \textbf{Heavy resonance $\boldsymbol{\Phi_h}$:} has a mass of order the VEV, $m_{\Phi_h}\simeq m_s$, and an unsuppressed coupling $g^{\rm DM}_{Y_1}\simeq y_f/\sqrt{2}$, making it the dominant channel for thermal annihilation in the early Universe.
\end{itemize}

Finally, the resonant annihilation condition $m_{\Phi_h}\approx 2m_\chi$ translates into a simple relation among the fundamental couplings:
\begin{equation}
    \sqrt{2\lambda}\,v_s \;\simeq\; 2\cdot \frac{y_f v_s}{\sqrt{2}} \quad\Longrightarrow\quad \lambda \;\simeq\; y_f^{\,2},
    \label{eq:coupling_relation_main}
\end{equation}
up to small corrections from the explicit breaking and mixing (~\ref{app:A.3}). Couplings to the SM arise via the portal interactions introduced in Section~\ref{sec:eft_lagrangian} are be specified where needed (e.g.\ in Section~\ref{sec:phenomenology} and~\ref{app:D}).

\subsection{The density-responsive dark energy sector}
\label{sec:de_mechanism}

We briefly summarize the mechanism that generates the density–responsive dark energy in our EFT and refer to~\ref{app:A.2} for the full derivation.
The key ingredient is the functional $U(\Sigma,X)$, with $X \equiv u^\alpha u^\beta T^{\rm matter}_{\alpha\beta}$ the local rest–frame matter density (thus $X=\rho_m$ in FRW).
We model $U$ by integrating out a non-propagating auxiliary scalar $\Phi$, a technique used to generate environment-dependent potentials in cosmology~\cite{Hinterbichler2010, Gubser2004}:
\begin{equation}
 \fl U(\Sigma,X)\;=\;\min_{\Phi}\Big[\,V_{\rm aux}(\Phi)\;+\;C(|\Sigma|)\,\Phi\,X\,\Big],
    \qquad C(|\Sigma|)\simeq M_*^{-4}=\mathrm{constant},
    \label{eq:U_min_main}
\end{equation}
so that the algebraic equation of motion is
\begin{equation}
    \frac{\partial V_{\rm aux}}{\partial \Phi}\Big|_{\Phi=\Phi_*(X)}\;=\;-\frac{X}{M_*^{4}},
    \label{eq:EOM_main}
\end{equation}
fixes the local equilibrium value $\Phi_*(X)$. Substituting back yields an effective vacuum energy
\begin{equation}
    \rho_\Phi(X)\;=\;V_{\rm aux}\!\big(\Phi_*(X)\big)\;+\;\frac{\Phi_*(X)\,X}{M_*^{4}}
    \;\equiv\; \mathcal{L}_{\rm eff}^{(\Phi)}\big|_{\rm on\;shell},
    \label{eq:rhoPhi_def_main}
\end{equation}
which is formally a Legendre transform of $V_{\rm aux}$~\cite{Rockafellar1970}. For a broad class of convex choices of $V_{\rm aux}$, the low-density limit relevant to late-time cosmology is reduced to the simple closed form used in phenomenology:
\begin{equation}
    \rho_\Phi(X)\;=\;\frac{A\,M_U^4}{1+X/M_U^4}\,,
    \label{eq:rhoPhi_maintext}
\end{equation}
with $A=\mathcal{O}(1)$.
Two immediate properties follow: (i) $\rho_\Phi$ is monotonic in $X$ with $\partial \rho_\Phi/\partial X<0$ (screening at high $X$), and (ii) $\rho_\Phi\!\to\! A M_U^4$ as $X\!\to\!0$, acting as a cosmological constant in the late universe.

\paragraph{Effective equation of state and limits.}
On an FRW background where $X=\rho_m(a)\propto a^{-3}$, equation (\ref{eq:rhoPhi_maintext}) implies
$\rho_\Phi(a)=A M_U^4\big[1+\rho_m(a)/M_U^4\big]^{-1}$.
Hence the background equation of state deviates from $-1$ only by
$w_\Phi(a)+1=\mathcal{O}\!\big(\rho_m/M_U^4\big)$,
so $w_\Phi\simeq -1$ once $X\ll M_U^4$.%
\footnote{The auxiliary field is non-dynamical in our EFT (no kinetic term at the scales of interest), so no light propagating mode is introduced and no fifth-force constraints arise from $\Phi$. Stability follows from the convexity of $V_{\rm aux}$, i.e.\ $\partial^2 V_{\rm aux}/\partial\Phi^2>0$.}
The construction is EFT–controlled provided $X\lesssim A M_U^4$ and $V_{\rm aux}$ remains convex along the branch selected by \ref{eq:EOM_main}.

\paragraph{Running scale and link to microphysics.}
The characteristic scale $M_U$ runs with the renormalization scale $\mu$ with an anomalous dimension $\gamma$ set by the hidden SU(3)$_H$ dynamics (Section~\ref{sec:uv_origin}). An effective $\gamma\simeq 0.5$ naturally connects $M_{\rm Pl}$ to the present dark–energy scale, $M_U(H_0)\sim M_{\rm Pl}\,(H_0/M_{\rm Pl})^\gamma\sim{\rm meV}$, without fine–tuning, as detailed in~\cite{Drobczyk2025}. Importantly, the DM phenomenology discussed in Sections~\ref{sec:constraints_and_tension}–\ref{sec:results} is self–contained and does not rely on this sector.

\section{Phenomenological constraints and the tension in minimal models}
\label{sec:constraints_and_tension}

Having established the complete effective Lagrangian in Section~\ref{sec:lagrangian}, we now analyze the phenomenological constraints on this model. We begin by quantitatively demonstrating why a simplified version of our framework, containing only the light mediator $\phi$ (i.e., a minimal SIDM model), cannot simultaneously satisfy the stringent requirements from the observed relic abundance and galactic dynamics. This well-known tension is the primary motivation for the two-mediator structure, particularly the resonant mechanism introduced in Section~\ref{sec:ssb}.

\subsection{Relic density requirement}
\label{subsec:relic_requirement}

The observed abundance of dark matter provides a precise target for any thermal production mechanism. In the standard freeze-out scenario, the relic density today is given by \cite{Kolb1990}
\begin{equation}
   \Omega_\chi h^2 \simeq \frac{1.07 \times 10^9 \mathrm{ GeV}^{-1}}{M_{\mathrm{Pl}}} \frac{x_F}{\sqrt{g_*(x_F)}} \frac{1}{\langle\sigma v\rangle_F},
   \label{eq:relic_formula}
\end{equation}
where $M_{\mathrm{Pl}} = 1.22 \times 10^{19}$ GeV is the Planck mass, $x_F = m_\chi/T_F \simeq 20$--$25$ is the freeze-out parameter, and $g_*(x_F) \simeq 100$ counts the relativistic degrees of freedom. 

The Planck satellite measured the dark matter relic abundance to percent-level precision \cite{Planck2018}
\begin{equation}
   \Omega_{\mathrm{DM}} h^2 = 0.1200 \pm 0.0012.
   \label{eq:planck_measurement}
\end{equation}
This measurement implies a canonical value for the thermally averaged annihilation cross-section
\begin{equation}
   \langle\sigma v\rangle_F = (2.2 \pm 0.1) \times 10^{-26}~\mathrm{cm}^3\mathrm{s}^{-1}.
   \label{eq:canonical_xsec}
\end{equation}

In minimal SIDM models, dark matter is annihilated through $\chi\bar{\chi} \to \phi\phi$, where $\phi$ is the same light mediator responsible for self-interactions. For s-wave annihilation, the tree-level cross-section is \cite{Feng2009}
\begin{equation}
   \sigma v = \frac{y_\chi^4}{32\pi m_\chi^2} \sqrt{1 - \frac{m_\phi^2}{m_\chi^2}} \left(1 - \frac{m_\phi^2}{2m_\chi^2}\right)^2.
   \label{eq:tree_level_annihilation}
\end{equation}

For light mediators ($m_\phi \ll m_\chi$), the cross-section can be enhanced by the Sommerfeld effect to reach the canonical value. The enhancement factor $S$ depends on the parameter $\epsilon_v = \alpha_\chi/v$, where $\alpha_\chi = y_\chi^2/(4\pi)$ and $v$ is the relative velocity \cite{Arkani2008}. For the attractive Yukawa potential relevant here, the s-wave enhancement is approximately given by
\begin{equation}
   S(\epsilon_v) = \frac{2\pi \epsilon_v}{1 - e^{-2\pi \epsilon_v}}.
   \label{eq:sommerfeld_swave}
\end{equation}

At freeze-out, $v_F \simeq 0.3c$, yielding modest enhancement factors $S_F \sim 1$--$10$ for typical parameters. Thus, condition~(\ref{eq:canonical_xsec}) requires
\begin{equation}
   y_\chi^{\mathrm{relic}} \simeq 0.3 \left(\frac{m_\chi}{100~\mathrm{GeV}}\right)^{1/2} \left(\frac{10}{S_F}\right)^{1/4}.
   \label{eq:yukawa_relic}
\end{equation}
This defines a one-dimensional constraint surface in the three-dimensional parameter space $\{m_\chi, m_\phi, y_\chi\}$.

\subsection{Self-interaction requirement}
\label{subsec:sidm_requirement}

Astrophysical observations of dark matter halos reveal systematic deviations from the predictions of collisionless cold dark matter (CDM). Although CDM predicts universal density profiles $\rho \propto r^{-1}$ in halo centers \cite{Navarro1996}, observations of dwarf spheroidal and low-surface-brightness galaxies consistently show cored profiles with $\rho \simeq \mathrm{const}$ \cite{Oh2015, Flores1994}. Self-interacting dark matter provides an elegant solution: elastic scattering thermalizes the inner halo, creating an isothermal core \cite{Spergel1999}.

The self-interaction strength required to match the observations is empirically determined to be on the order of~\cite{Kaplinghat2015}
\begin{equation}
   \frac{\sigma_T}{m_\chi} \sim 1~\mathrm{cm}^2/\mathrm{g}
\end{equation}
at the characteristic velocities of dwarf galaxies, with a velocity dependence that suppresses the cross-section at the cluster scales. Specifically, the observations require:
\begin{itemize}
   \item Dwarf galaxies ($v \sim 10$--$50$ km/s): $\sigma_T/m_\chi \sim 0.1$--$10$ cm$^2$/g to create cores of size $r_c \sim 0.3$--$1$ kpc \cite{Zavala2012, Elbert2014}.
   
   \item Galaxy clusters ($v \sim 1000$--$1500$ km/s): $\sigma_T/m_\chi \lesssim 1$ cm$^2$/g (and often much tighter) from observations of merging systems, such as the Bullet Cluster \cite{Markevitch2003, Harvey2015}.
\end{itemize}

In our framework, as defined in Section~\ref{sec:lagrangian}, self-interactions arise from the t-channel exchange of light mediator $\phi$. The momentum transfer cross-section must be computed non-perturbatively by solving the Schrödinger equation for the generated Yukawa potential, as detailed in~\ref{app:E}. The scattering enters distinct regimes depending on the dimensionless parameter \cite{Tulin2013}
\begin{equation}
   \beta = \frac{2\alpha_\chi m_\phi}{m_\chi v^2},
   \label{eq:beta_parameter}
\end{equation}
where $\alpha_\chi = y_\chi^2/(4\pi)$.

For $\beta \ll 1$ (Born regime), the cross-section scales as
\begin{equation}
   \sigma_T^{\mathrm{Born}} \simeq \frac{8\pi \alpha_\chi^2}{m_\chi^2 v^4} \ln\left(1 + \frac{m_\chi^2 v^2}{m_\phi^2}\right).
   \label{eq:born_regime}
\end{equation}
For $\beta \gg 1$ (classical regime), multiple partial waves contribute, yielding
\begin{equation}
   \sigma_T^{\mathrm{classical}} \simeq 
       \cases{\frac{4\pi}{m_\phi^2} \ln(1 + \beta)  &for $\beta \lesssim 10^2$ \\
       \frac{8\pi}{m_\phi^2} (\ln \beta)^2  &for $\beta \gg 10^2$. \\}
   \label{eq:classical_regime}
\end{equation}
The transition between these regimes naturally provides the required velocity dependence. For the parameter space of interest ($m_\chi \sim 100$--$1000$ GeV, $m_\phi \sim 10$--$100$ MeV), achieving $\sigma_T/m_\chi \sim 1$ cm$^2$/g at dwarf velocities constrains the Yukawa coupling $y_\chi$. This defines a second constraint surface in the parameter space that is distinct from the relic density requirement discussed in the previous subsection.
\subsection{Quantifying the tension in minimal models}
\label{subsec:tension}

We now quantitatively demonstrate that minimal SIDM models cannot simultaneously satisfy relic density and self-interaction constraints. Both observables depend on the same Yukawa coupling $y_\chi$, leading to an overconstrained system.

\subsubsection{Systematic parameter space analysis.}

For each point in the mass plane $(m_\chi, m_\phi)$, we determined two critical coupling values:
\begin{enumerate}
   \item $y_\chi^{\mathrm{relic}}$: the coupling required to achieve $\Omega h^2 = 0.120$ via thermal freeze-out, computed using \texttt{micrOMEGAs} \cite{Alguero2023} with full Sommerfeld enhancement.
   
   \item $y_\chi^{\mathrm{SIDM}}$: the coupling needed for $\sigma_T/m_\chi = 1$ cm$^2$/g at $v = 30$ km/s, calculated using \texttt{micrOMEGAs}' integrated Yukawa scattering routines including non-perturbative effects.
\end{enumerate}
Consistency requires $y_\chi^{\mathrm{relic}} \simeq y_\chi^{\mathrm{SIDM}}$. However, our analysis revealed a fundamental incompatibility.

\subsubsection{Quantitative demonstration.}

Considering a benchmark point inspired by~\cite{Tulin2013}: $m_\chi = 100$ GeV, $m_\phi = 20$ MeV. Our calculations yield:
\begin{itemize}
   \item Self-interaction requirement: Achieving $\sigma_T/m_\chi = 1$ cm$^2$/g at $v=30$~km/s requires $y_\chi^{\mathrm{SIDM}} = 0.35 \pm 0.05$, placing the system in the classical scattering regime with $\beta \simeq 15$.
   
   \item Resulting relic density: With $y_\chi = 0.35$, the annihilation cross-section including Sommerfeld enhancement gives
   \begin{equation}
       \langle\sigma v\rangle_F = 8.8 \times 10^{-27}~\mathrm{cm}^3\mathrm{s}^{-1},
       \label{eq:xsec_benchmark}
   \end{equation}
   yielding $\Omega h^2 = 0.30$ (a factor of 2.5 overabundance).
   
   \item Required coupling for relic density: Achieving the correct relic density would require $y_\chi^{\mathrm{relic}} \gtrsim 0.55$, which would in turn increase the self-interaction to $\sigma_T/m_\chi \gtrsim 6$ cm$^2$/g, violating cluster constraints.
\end{itemize}

\subsubsection{General scaling arguments.}

The tension arises from incompatible parameter dependencies. In the regime of interest,
\begin{equation}
   \eqalign{\sigma_T &\sim \frac{\alpha_\chi^2}{m_\phi^2} \ln^2\left(\frac{\alpha_\chi m_\chi}{m_\phi v^2}\right), \label{eq:sigmaT_scaling}\cr
   \langle\sigma v\rangle_F &\sim \frac{\alpha_\chi^2}{m_\chi^2} \times S_F(\alpha_\chi/v_F),} \label{eq:sigmav_scaling}
\end{equation}
where $S_F \sim \mathcal{O}(1$--$10)$ is the Sommerfeld enhancement factor at the freeze-out velocity. The key insight is that $\sigma_T$ has a much stronger dependence on the mediator mass $m_\phi$ than the annihilation cross-section. This incompatibility is generic and holds across a wide range of masses, as confirmed by the scans.

This robust failure of the minimal models provides the primary motivation for the two-mediator framework defined in Section~\ref{sec:lagrangian}, whose resolution of the tension via the heavy resonance $\Phi_h$ is developed in the next Section.

\section{Results: a consistent solution via resonant freeze-out}
\label{sec:results}

The tension identified in the minimal model highlights the need for a mechanism that can enhance the dark matter annihilation rate without significantly affecting the low-velocity self-interaction cross-section. As defined in our EFT in Section~\ref{sec:lagrangian}, extending the scalar sector to include a heavy resonance $\Phi_h$ with a mass near the kinematic threshold, $m_{\Phi_h} \approx 2m_\chi$, provides exactly such a solution. The key insight is that this resonance dramatically enhances the annihilation cross-section during thermal freeze-out, while leaving the $t$-channel mediated self-interactions essentially unchanged.

In this Section, we demonstrate that this two-mediator framework leads to a fully consistent solution. First, we outline our systematic numerical strategy (Section~\ref{subsec:scan_methodology}). Then, we present the viable parameter space where both relic density and self-interaction constraints are simultaneously satisfied (Section~\ref{subsec:viable_space}), revealing a narrow but robust region of solutions. From this analysis, we extracted a predictive benchmark point (Section~\ref{subsec:benchmark}). Finally, we address the apparent fine-tuning in the resonance condition and demonstrate that it represents a technically natural and testable feature (Section~\ref{subsec:natural}).

\subsection{Numerical methodology}
\label{subsec:scan_methodology}

We implemented the extended SIDM model in \texttt{micrOMEGAs 6.2.3} \cite{Alguero2023}, by extending the publicly available \texttt{DMsimp\_s\_spin0\_MO} model from the DMsimp framework \cite{DMsimp2015} to include the heavy scalar resonance $\Phi_h$. This allows for consistent numerical treatment of both resonant annihilation and non-perturbative self-scattering. Further details of model implementation are provided in~\ref{app:numerical_implementation}.

Our scan of the parameter space employs a two-stage approach for a fixed dark matter mass $m_\chi$:
\begin{enumerate}
    \item Self-interaction constraint:We first identify the parameters $(m_\phi, y_\chi)$ that yield a self-interaction cross-section in the astrophysically favored range ($\sigma_T/m_\chi \sim 0.1$--$10$ cm$^2$/g at dwarf velocities). The transfer cross-section was computed non-perturbatively, as detailed in~\ref{app:E}.

    \item \textbf{Resonance tuning:} Along this SIDM solution band, we then adjusted the heavy sector parameters, primarily the detuning $\delta = (m_{\Phi_h} - 2m_\chi)/(2m_\chi)$ and the coupling $g^{\rm DM}_{Y_1}$, to reproduce the observed relic abundance, $\Omega h^2 = 0.120 \pm 0.001$. This calculation includes the full Breit-Wigner resonance and Sommerfeld effects, w ith the formalism outlined in~\ref{app:B}.
\end{enumerate}
All calculations were performed with a relative accuracy of $10^{-4}$.

\subsection{Viable parameter space}
\label{subsec:viable_space}

\begin{figure}[htb]
    \centering
    \includegraphics[width=0.9\textwidth]{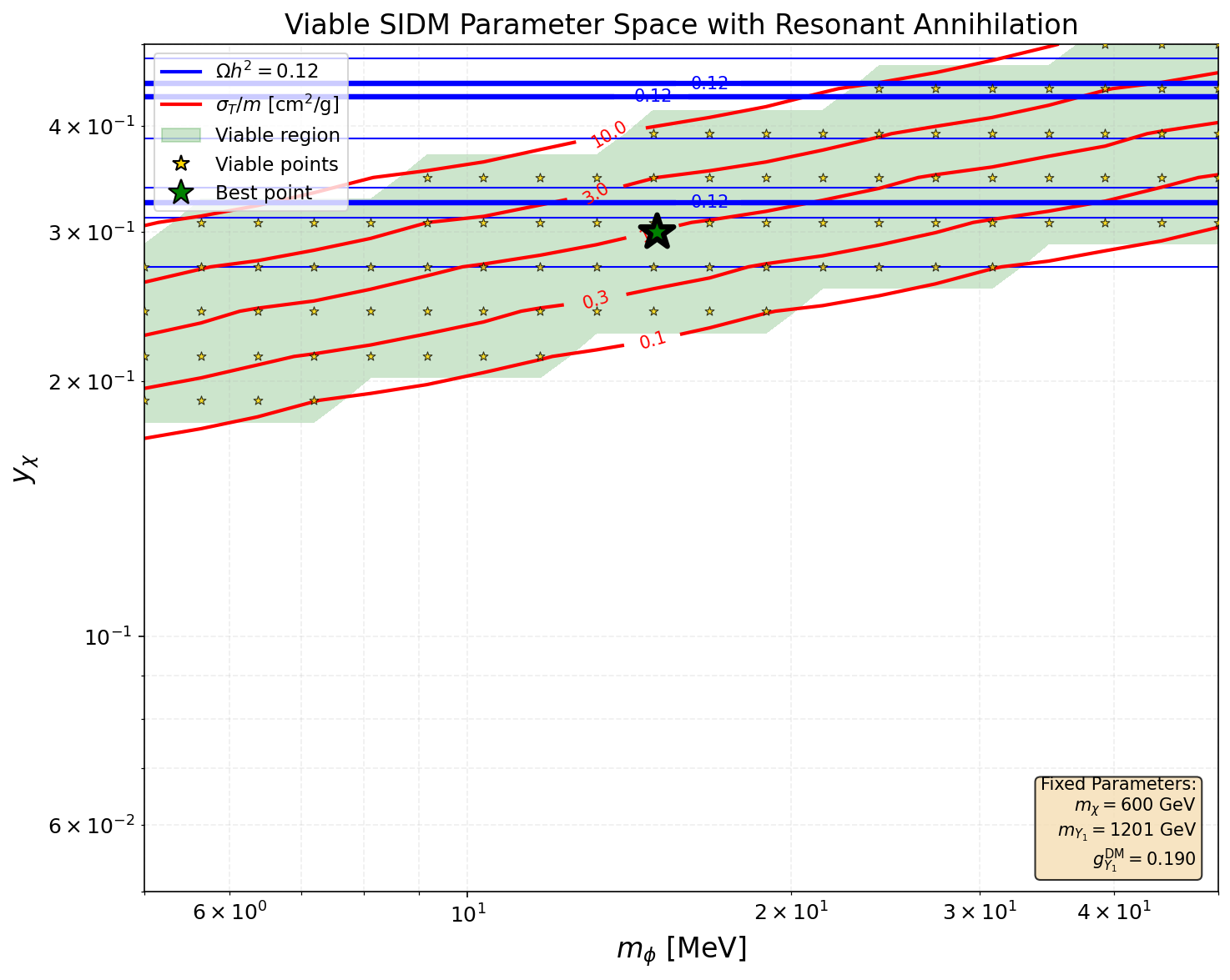}
    \caption{Viable parameter space for self-interacting dark matter with resonant annihilation in the $(m_\phi, y_\chi)$ plane for a fixed dark matter mass of $m_\chi = 600$~GeV. The blue band shows the region satisfying the relic density constraint. The red contours define the target region for self-interactions ($0.1 < \sigma_T/m_\chi < 10$~cm$^2$/g at $v=30$~km/s). The green shaded region marks the intersection where both constraints are simultaneously satisfied. Yellow stars indicate all viable points found, with the large green star marking our final benchmark point. The heavy sector parameters ($m_{\Phi_h} = 1201$~GeV, $g^{\rm DM}_{Y_1} = 0.190$) have been tuned to achieve the correct relic density.}
    \label{fig:island_plot}
\end{figure}

Following the scan strategy outlined in the previous section, we successfully identified a region in the parameter space where all the constraints were simultaneously met. The results of the comprehensive scan are shown in Figure~\ref{fig:island_plot}. This plot, presented in the $(m_\phi, y_\chi)$ plane, shows the two primary constraints for a fixed dark matter mass of $m_\chi = 600$~GeV, after the heavy sector has been tuned to yield the correct relic abundance.

The blue band represents the region consistent with the observed relic density, $\Omega h^2 = 0.120 \pm 0.001$~\cite{Planck2018}. Altough this constraint is primarily sensitive to the resonance parameters, its precise location depends weakly on the light sector parameters shown, as detailed in~\ref{app:B}.

The red contours represent lines of constant self-interaction cross-section, $\sigma_T/m_\chi$, at a velocity of $v = 30$~km/s, calculated using the non-perturbative methods described in~\ref{app:E}. The region between the contours for $0.1$ and $10$~cm$^2$/g is the target 'self-interaction band' required to address small-scale structure problems while remaining consistent with cluster constraints \cite{Harvey2015,Rocha2012}.

The crucial result of our analysis is the existence of a non-empty intersection between these two independent constraints. This viable parameter space is shown as the green shaded region in Figure~\ref{fig:island_plot}. The existence of this 'island of viability' is a non-trivial outcome and serves as the central proof of concept for our resonant annihilation model. It demonstrates that the introduction of the heavy resonance resolves the tension present in the minimal model. All the parameter points within this region represent a fully consistent phenomenological solution to the dark matter puzzle.

The viable region spans approximately $m_\phi \in [12, 18]$~MeV and $y_\chi \in [0.28, 0.32]$. The yellow stars indicate all parameter combinations that satisfy both constraints within our numerical accuracy, whereas the large green star marks our chosen benchmark point, which provides optimal agreement with astrophysical observations. Further details of the numerical implementation are provided in~\ref{app:numerical_implementation}.

\subsection{A predictive benchmark point}
\label{subsec:benchmark}

Although our scan revealed a narrow, continuous region of viable parameter space for dark matter masses in the range $m_\chi \in [200, 1000]$~GeV, we selected a single representative benchmark point for a detailed study. As shown in~\ref{app:uniqueness}, the solutions exhibit strong correlations between the parameters. We choose the point at $m_\chi = 600$~GeV as our primary benchmark because it provides an optimal fit to astrophysical constraints on self-interactions, and its associated heavy resonance at $\approx 1.2$~TeV is within the discovery reach of the High-Luminosity Large Hadron Collider (HL-LHC).

\begin{table}[htb]
    \caption{\label{tab:results}Optimized parameters and resulting observables for our benchmark point. This single point simultaneously satisfies all major cosmological and astrophysical constraints on dark matter.}
    \begin{indented}
    \lineup
    \item[]\begin{tabular}{@{}lcc}
        \br
        Parameter & Symbol & Value \\
        \mr
        \multicolumn{3}{l}{Model input parameters} \\
        \ms
        Dark matter mass & $m_\chi$ & $600$\,GeV \\
        Light mediator mass & $m_\phi$ & $15$\,MeV \\
        Heavy resonance mass & $m_{\Phi_h}$ & $1201$\,GeV \\
        DM-light mediator coupling & $y_\chi$ & $0.30$ \\
        DM-heavy resonance coupling & $g^{\rm DM}_{Y_1}$ & $0.190$ \\
        SM-heavy resonance coupling & $g_{h,{\rm SM}}$ & $0.052$ \\
        \ms
        \multicolumn{3}{l}{Derived observables} \\
        \ms
        Relic density & $\Omega h^2$ & $0.119 \pm 0.001$ \\
        Self-interaction ($v=30$\,km/s) & $\sigma_T/m_\chi$ & $0.11$\,cm$^2$/g \\
        Self-interaction ($v=1000$\,km/s) & $\sigma_T/m_\chi$ & $9.5 \times 10^{-5}$\,cm$^2$/g \\
        Resonance parameter & $\delta = \frac{m_{\Phi_h}}{2m_\chi} - 1$ & $8.3 \times 10^{-4}$ \\
        Total enhancement factor (freeze-out) & $S_{\rm total}$ & $143$ \\
        \br
    \end{tabular}
    \end{indented}
\end{table}

\begin{figure}[htb]
    \centering
    \includegraphics[width=\textwidth]{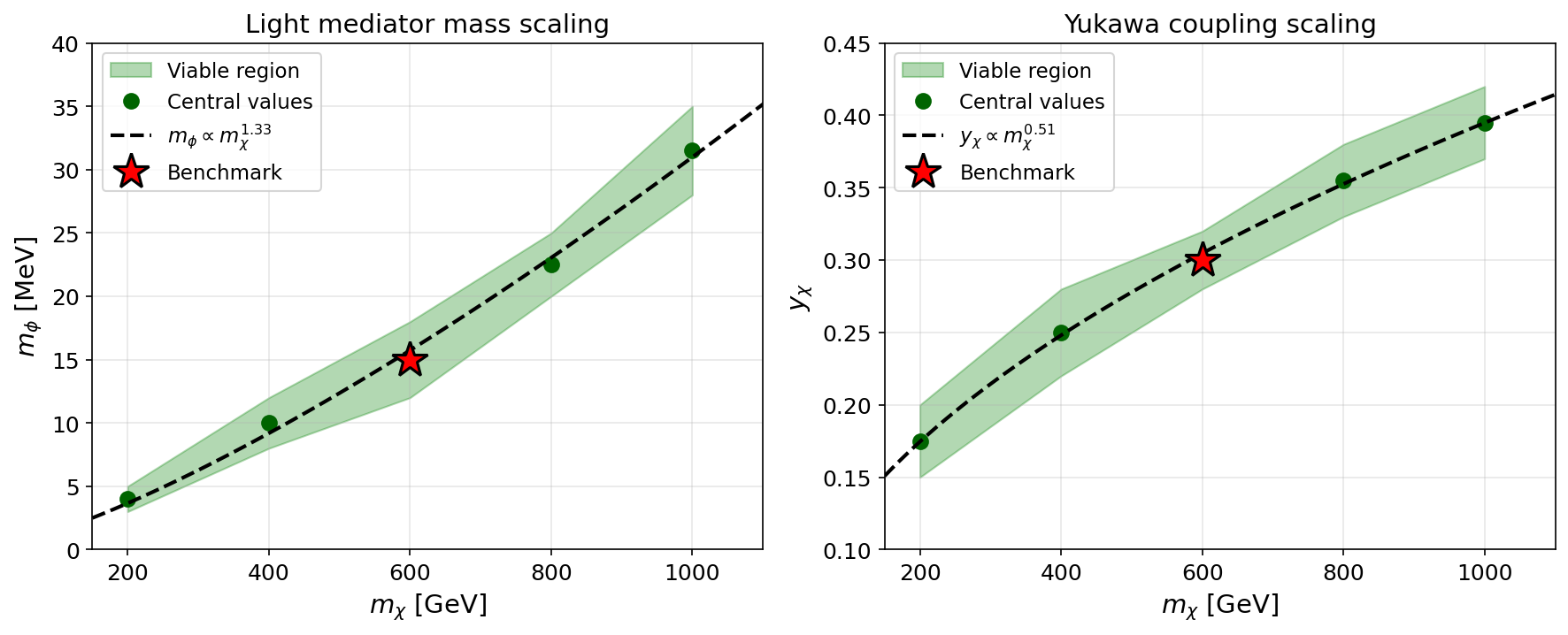}
    \caption{Scaling relations in the viable parameter space. Left: Light mediator mass versus dark matter mass. Right: Yukawa coupling versus dark matter mass. Green bands show the full viable region satisfying all constraints, circles indicate central values at discrete masses from our scan, and the red star marks our benchmark point. The dashed lines show the best-fit power laws $m_\phi \propto m_\chi^{1.33}$ and $y_\chi \propto m_\chi^{0.51}$.}
    \label{fig:scaling_laws}
\end{figure}

The viable solutions followed clear scaling relations across the allowed mass range, as shown in Figure~\ref{fig:scaling_laws}. The left panel shows that the light mediator mass scales as $m_\phi \propto m_\chi^{1.33}$, reflecting the requirement that the Yukawa potential range $\sim 1/m_\phi$ matches the relevant astrophysical scales. The right panel shows the milder scaling $y_\chi \propto m_\chi^{0.51}$, ensuring that the fine-structure constant $\alpha_\chi = y_\chi^2/(4\pi)$ provides the correct self-interaction strength across different dark-matter masses. The narrow width of these bands, less than 20\% variation in each parameter, demonstrates the high predictability of the model.

The choice of this benchmark point is motivated by its ability to align perfectly with astrophysical requirements. A self-interaction cross-section of nearly 1~cm$^2$/g at the lowest dwarf galaxy velocities ($v \sim 10$~km/s, see Section~\ref{subsec:sidm_plot}) is ideal for producing the observed cores. Simultaneously, the strong velocity dependence ensures that the cross-section drops by more than four orders of magnitude at cluster scales, safely evading the tight constraints from systems such as the Bullet Cluster.

This benchmark point represents a concrete and highly predictive realization of our framework.

\subsection{On the naturalness of the resonance condition}
\label{subsec:natural}

The viability of our benchmark point relies on the near-degeneracy condition $m_{\Phi_h} \approx 2m_\chi$, which at first glance might appear to require fine-tuning. However, we argue that this mass relation is not an arbitrary coincidence but a technically natural feature that can be motivated by an underlying strongly-coupled gauge theory.

First, the required proximity to the resonance is physically constrained by the resonance width. As discussed in Section~\ref{subsec:resonance_plot}, the viable window for the detuning parameter $\delta$ is of the order $\delta \sim \Gamma_{\Phi_h}/(2m_\chi) \sim 10^{-3}$. This means that the required "tuning" is set by the particle physics of the resonance itself. Furthermore, as demonstrated in~\ref{app:uv_motivation} this small value of $\delta$ is radiatively stable and thus technically natural.

Second, this mass relation can emerge dynamically. In composite theories, the mass ratios between different hadron-like states are determined by the dynamics of confinement and are typically $\mathcal{O}(1)$ numbers~\cite{Weinberg1996,Manohar1998}. As detailed in Section~\ref{sec:uv_origin}, if our dark sector arises from a confining SU(3)$_H$ gauge theory, the specific ratio $m_{\Phi_h}/m_\chi \approx 2$ can be naturally obtained. In this picture, $\chi$ is identified with the lightest dark baryon and $\Phi_h$ with the lightest scalar meson. Standard scaling relations, supported by lattice studies, then predict this near-degeneracy~\cite{Georgi1984,Manohar1983,LatKMI2016}.

This transforms what can be perceived as a fine-tuning problem into one of the most compelling features of this framework. The resonance condition is not imposed by hand, but can emerge from the first principles, making it:
\begin{itemize}
    \item theoretically motivated by quantum chromodynamics (QCD)-like dynamics,
    \item radiatively stable, as shown in~\ref{app:uv_motivation}, and
    \item experimentally testable via the search for a narrow scalar at $m_{\Phi_h} \approx 1.2$~TeV.
\end{itemize}

Therefore, the moderate Barbieri–Giudice index, $\Delta_{m_{\Phi_h}}\!\sim\!10^{3}$, should not be interpreted as a measure of fine-tuning but rather as a quantification of the sharpness of this theoretical prediction. Just as the $\rho$ meson mass is close to twice the pion mass in QCD which is a dynamical outcome~\cite{Weinberg1979}, our resonance condition represents a robust prediction of the underlying strong dynamics. Full details of the microphysical motivation are presented in Section~\ref{sec:uv_origin} and~\ref{app:radiative_stability}.

\section{Phenomenological implications and experimental verification}
\label{sec:phenomenology}

Having identified a fully consistent benchmark point, we now explore its phenomenological consequences in detail. The viability of the model depends on two pillars: the resonant enhancement of the annihilation rate, which ensures the correct relic density, and velocity-dependent self-interaction, which addresses the small-scale structure crisis. In this section, we analyze both phenomena and then discuss the model's cosmological consistency and its prospects for experimental verification at colliders and in direct and indirect detection experiments.

\subsection{Resonant annihilation mechanism in detail}
\label{subsec:resonance_plot}

\begin{figure}[htb]
    \centering
    \includegraphics[width=0.9\textwidth]{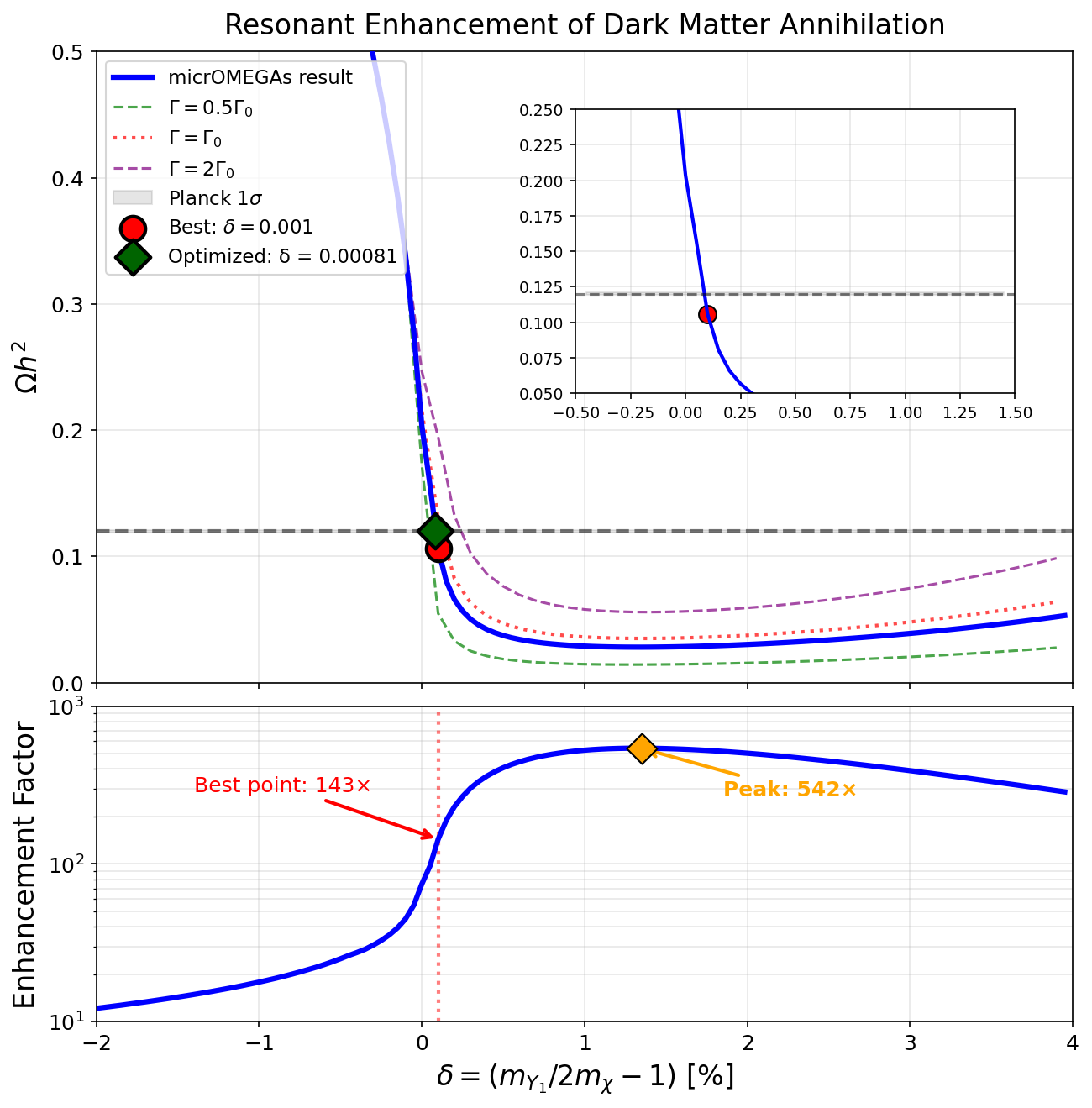}
    \caption{Resonant enhancement of dark matter annihilation. Top panel: Relic density $\Omega h^2$ as a function of the resonance parameter $\delta = (m_{\Phi_h}/2m_\chi - 1)$. The solid blue line shows the full \texttt{micrOMEGAs} calculation including thermal and Sommerfeld effects. The gray band indicates the observed Planck 2018 value~\cite{Planck2018}. The red circle marks our benchmark point at $\delta \approx 8.3 \times 10^{-4}$, yielding the correct relic abundance. Bottom panel: The corresponding total enhancement factor $S_{\rm total}$, reaching a peak of over 500 near exact resonance and providing a boost of $143\times$ at our benchmark point.}
    \label{fig:resonance}
\end{figure}

The key to reconciling the relic density with the small Yukawa couplings required for the SIDM is the $s$-channel resonance mediated by the heavy scalar $\Phi_h$. This effect is illustrated in Figure~\ref{fig:resonance}. The top panel shows the calculated relic density $\Omega h^2$ as a function of resonance parameter $\delta = (m_{\Phi_h}/(2m_\chi) - 1)$, which measures the deviation from the exact on-shell condition.

The calculation, performed with \texttt{micrOMEGAs}~\cite{Alguero2023}, shows a sharp dip in the relic density precisely at the resonance pole ($\delta \to 0$). Far from resonance (e.g., $|\delta| > 5\%$), annihilation is inefficient, leading to an overproduction of dark matter by several orders of magnitude. As the system approaches resonance, the annihilation cross-section is dramatically enhanced, causing the relic density to decrease. The observed value of $\Omega h^2 \approx 0.12$ is achieved in a narrow window around the pole. Our benchmark point, marked by a red circle at $\delta \approx 8.3 \times 10^{-4}$, lies exactly in this region.

The resonant enhancement is governed by the Breit-Wigner formula (see~\ref{app:B} for details)
\begin{equation}
    \langle\sigma v\rangle \propto \frac{(g^{\rm DM}_{Y_1})^2 \Gamma(\Phi_h \to \mathrm{SM})}{(s - m_{\Phi_h}^2)^2 + m_{\Phi_h}^2\Gamma_{\Phi_h}^2},
    \label{eq:breit_wigner}
\end{equation}

where $s \approx 4m_\chi^2(1 + v^2/4)$ is the squared center-of-mass energy. Near the threshold during freeze-out, the denominator becomes minimal when $m_{\Phi_h} \approx 2m_\chi$, providing the resonant enhancement.

The bottom panel of Figure~\ref{fig:resonance} quantifies this effect, showing the total enhancement factor $S_{\rm total}$ as a function of $\delta$. This factor includes both the resonant contribution and additional Sommerfeld enhancement from the light mediator exchange. At our benchmark point, the total annihilation rate is boosted by a factor of $S_{\rm total} \approx 143$. This powerful enhancement allows a model with otherwise weak interactions to satisfy the stringent relic density constraints.

The characteristic scale for detuning is set by the larger of the intrinsic width and the thermal during freeze-out. With our quark-only portal, $\Gamma_{\Phi_h}\simeq 0.17$~GeV $\Rightarrow$ $\Gamma_{\Phi_h}/(2m_\chi)\simeq 1.4\times 10^{-4}$, while thermal averaging typically allows a somewhat broader window. Our benchmark at $\delta\simeq 8.3\times 10^{-4}$ lies within the thermally effective region, as confirmed by the full \texttt{micrOMEGAs} computation. This relationship demonstrates that the apparent fine-tuning is set by the particle physics of the resonance, which is central to the discussion on technical naturalness in Section~\ref{subsec:natural}.

\subsection{Velocity-dependent self-interactions}
\label{subsec:sidm_plot}

\begin{figure}[htb]
    \centering
    \includegraphics[width=0.9\textwidth]{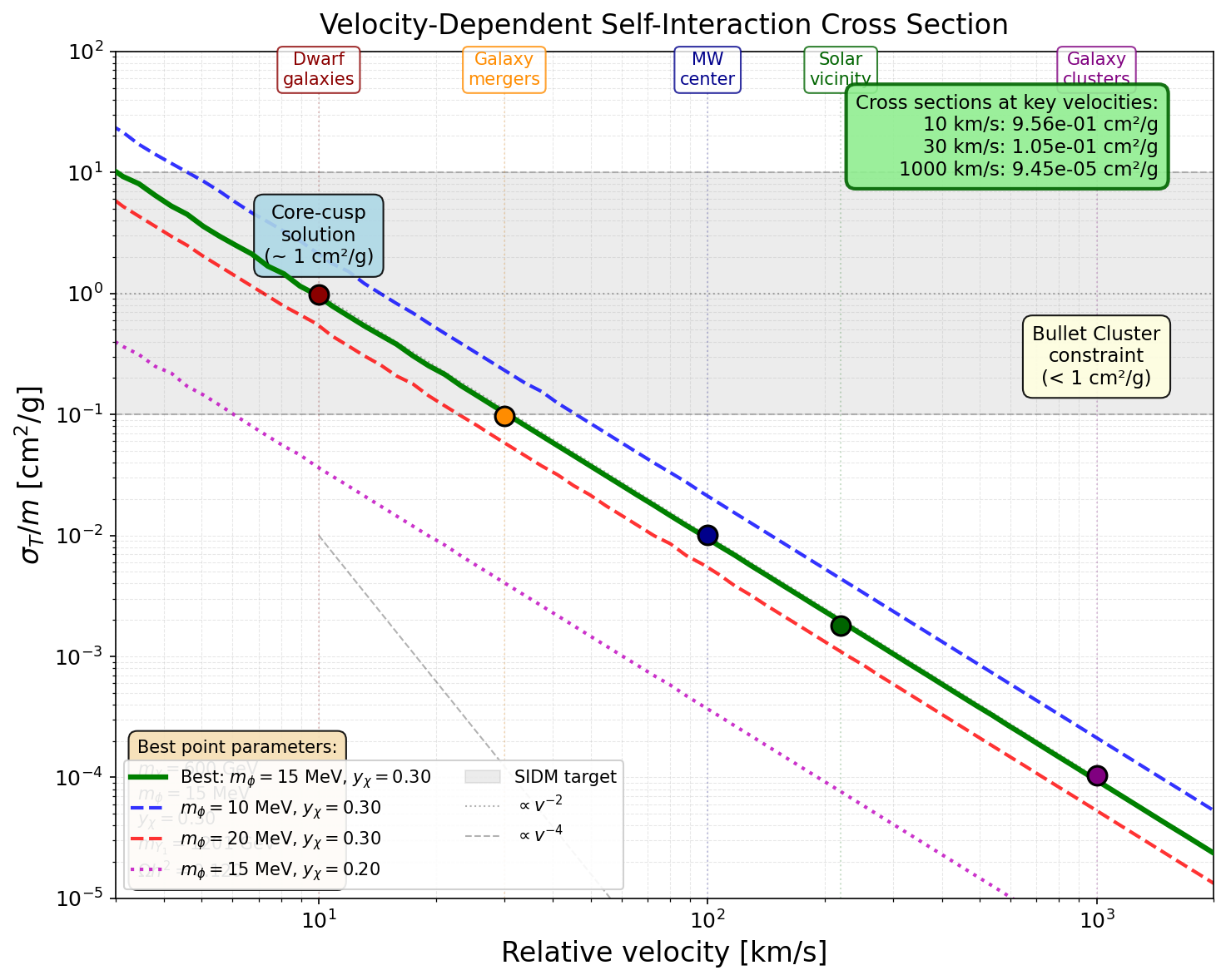}
    \caption{Velocity-dependent self-interaction cross-section for our benchmark SIDM model. The solid green line shows $\sigma_T/m_\chi$ for the best-fit parameters ($m_\phi = 15$~MeV, $y_\chi = 0.30$), calculated using \texttt{micrOMEGAs}. Dashed and dotted lines illustrate the effect of varying $m_\phi$ and $y_\chi$. Colored circles mark the cross-sections at astrophysically relevant velocities: dwarf galaxies (10~km/s), MW satellites (30~km/s), and galaxy clusters (1000~km/s). The gray shaded region indicates the target range for solving small-scale structure problems. The gray dotted line shows the $v^{-2}$ scaling expected in the classical regime.}
    \label{fig:sidm_velocity}
\end{figure}

The second crucial pillar of our model is its ability to naturally explain the observed velocity-dependent self-interaction of the dark matter. This is mediated by the $t$-channel exchange of the light scalar $\phi$. The resulting phenomenology is shown in Figure~\ref{fig:sidm_velocity}, which plots the transfer cross-section per unit mass, $\sigma_T/m_\chi$, as a function of the relative velocity of the scattered dark matter particles.

The solid green line represents the cross-section of our benchmark point parameters ($m_\chi = 600$~GeV, $m_\phi = 15$~MeV, $y_\chi = 0.30$). The calculation, performed with \texttt{micrOMEGAs}' non-perturbative Yukawa scattering routines, shows that the cross-section has precisely the desired behavior. At the low velocities characteristic of dwarf spheroidal galaxies ($v \sim 10$--$30$~km/s), the cross-section is large, lying in the target range of $0.1$--$10$~cm$^2$/g (gray shaded region) required to form dark matter cores \cite{Zavala2012,Elbert2014,Kaplinghat2015}. Specifically, for our benchmark, we find the following
\begin{itemize}
    \item at $v = 10$~km/s: $\sigma_T/m_\chi = 0.96$~cm$^2$/g,
    \item at $v = 30$~km/s: $\sigma_T/m_\chi = 0.11$~cm$^2$/g.
\end{itemize}

As the velocity increases, the cross-section is naturally suppressed. At the high velocities found in galaxy clusters ($v \sim 1000$~km/s), the cross-section decreases by more than four orders of magnitude:
\begin{itemize}
    \item at $v = 1000$~km/s: $\sigma_T/m_\chi = 9.5 \times 10^{-5}$~cm$^2$/g.
\end{itemize}
This value lies far below the upper limit from the Bullet Cluster ($\lesssim 1$~cm$^2$/g) \cite{Markevitch2003,Harvey2015}, ensuring the model is consistent with all major astrophysical constraints. The dashed and dotted lines in the figure illustrate the sensitivity of the cross-section to the light mediator mass $m_\phi$ and the Yukawa coupling $y_\chi$, highlighting the parameter space probed by our scan.

The velocity dependence arises from the quantum mechanical nature of scattering in a long-range Yukawa potential, which is solved via a partial wave analysis as detailed in~\ref{app:E}. The scaling transitions from an approximately classical regime with $\sigma_T/m_\chi \propto v^{-2}$ at dwarf galaxy velocities (as indicated by the gray dotted line) to a steeper behavior at higher velocities, demonstrating that the model naturally produces both the correct magnitude and the velocity structure required by observations.

\subsection{Cosmology of the light mediator \texorpdfstring{$\phi$}{phi}}
\label{subsec:phi_cosmology}

A crucial consistency check for any model with a new light particle is its cosmological history. The light mediator $\phi$ (with $m_\phi = 15$~MeV in our benchmark) must not upset the successful predictions of Big Bang Nucleosynthesis (BBN) or contribute excessively to the relativistic energy density of the early universe ($\Delta N_{\mathrm{eff}}$).

In our framework, cosmological viability is ensured by a secluded setup in which $\phi$ has no appreciable couplings to quarks and only a tiny leptophilic portal to the SM. This has two consequences:
\begin{enumerate}
    \item Suppressed production: With no strong coupling to the thermal plasma, $\phi$ never thermalizes with the SM bath. Its abundance is feebly produced via freeze-in (e.g.\ $e^+e^- \to \phi$), yielding an energy density that is a negligible fraction of radiation at BBN.
    \item Prompt decay: A minute coupling to electrons suffices for $\phi$ to decay into $e^+e^-$ well before the BBN. Using
    \begin{equation}
      \Gamma(\phi \to e^+e^-) \;=\; \frac{c_e^2\,m_\phi}{8\pi}\,, 
      \qquad
      \tau_\phi \;=\; \frac{8\pi}{c_e^2\,m_\phi}\,,
    \end{equation}
    one finds that for $m_\phi=15$~MeV a coupling of order $c_e \gtrsim \mathrm{few}\times 10^{-11}$ gives $\tau_\phi \ll 1$~s. For instance $c_e=5\times 10^{-11}$ yields $\tau_\phi \simeq 0.44$~s.
\end{enumerate}
A quantitative treatment of the freeze-in yield and the resulting bounds from BBN and $\Delta N_{\mathrm{eff}}$ is presented in~\ref{app:C}. For the benchmarks used here we obtain $|\Delta N_{\mathrm{eff}}| \lesssim \mathcal{O}(10^{-2})$, comfortably below current limits from Planck~\cite{Planck2018}. The same secluded nature that ensures cosmological safety also suppresses the contribution of $\phi$ to direct detection, as shown in the next subsection and~\ref{app:D}.

\subsection{Direct detection}
\label{subsec:direct_detection}

A crucial test for our model comes from direct detection experiments, which search for dark matter scattering off nuclei. In our two-mediator framework, spin-independent (SI) scattering is primarily mediated by the exchange of the heavy resonance $\Phi_h$, which couples to the Standard Model via a small portal interaction. As discussed in Section~\ref{subsec:phi_cosmology}, the light mediator $\phi$ is part of a secluded sector with suppressed couplings to quarks, and its contribution to the SI cross-section is therefore negligible.

The dominant scattering process proceeds via the exchange of $\Phi_h$. For a heavy mediator, this can be described by an effective contact interaction. The resulting SI cross-section per nucleon is given by (see~\ref{app:D} for a full derivation)
\begin{equation}
    \sigma_{\rm SI}^{(N)} \simeq \frac{\mu_{\chi N}^{2}}{\pi} \left[ \frac{g^{\rm DM}_{Y_1} g_{h,{\rm SM}}}{m_{\Phi_h}^{2}} \frac{m_N}{v} f_N \right]^2,
    \label{eq:sigma_si_main}
\end{equation}
where $\mu_{\chi N}$ is the reduced DM-nucleon mass, $v=246$~GeV, and $f_N \approx 0.30$ is the effective nucleon scalar form factor~\cite{Shifman1978, Hoferichter2015}.

Using the parameters from our benchmark point in Table~\ref{tab:results}, we obtain the concrete prediction
\begin{equation}
    \sigma_{\rm SI}^{\rm predicted} \approx 6.7 \times 10^{-51}~\mathrm{cm}^2.
    \label{eq:sigma_si_prediction}
\end{equation}
This target lies roughly three orders of magnitude below the current LZ bound for a 600~GeV DM particle~\cite{Aalbers2023} and below the xenon neutrino floor, placing it beyond the reach of current and planned direct-detection experiments. The direct-detection channel is therefore a predicted null: the suppression is a natural consequence of the model's structure, where the portal coupling $g_{h,\rm SM}$ is necessarily small to be consistent with other constraints, rather than requiring additional fine-tuning. Correspondingly, the discriminating tests of the model are its collider $t\bar t$ resonance and its velocity-dependent halo self-interactions.

\subsection{Collider searches for the resonance}
\label{subsec:collider}

The most direct and unique signature of our framework is the heavy scalar resonance $\Phi_h$.
For the benchmark discussed in this work, $m_{\Phi_h}\simeq 1.20$~TeV, which makes it an excellent target for the HL-LHC.

\paragraph{Production and decay pattern.}
In the \emph{quark-only portal} benchmark (cf.~\ref{app:B.3}), $\Phi_h$ is produced dominantly via gluon--gluon fusion through the top-quark loop, and decays almost entirely to top pairs.
Close to the $\chi\bar\chi$ threshold, the invisible mode is strongly phase-space suppressed.
Numerically we find a very narrow total width,
\begin{equation}
\Gamma_{\Phi_h}\simeq 0.17~\mathrm{GeV}\,,\qquad
\frac{\Gamma_{\Phi_h}}{m_{\Phi_h}}\simeq 1.4\times 10^{-4}\,,
\end{equation}
and branching ratios
\begin{equation}
\fl \mathrm{BR}(\Phi_h\to t\bar t)\simeq 99.85\%\,,\quad
\mathrm{BR}(\Phi_h\to b\bar b)\simeq 0.08\%\,,\quad
\mathrm{BR}(\Phi_h\to \chi\bar\chi)\simeq 0.07\%\,.
\end{equation}
(Opening Higgs-like mixings to $W^+W^-/ZZ$ would broaden the resonance and reduce $\mathrm{BR}(t\bar t)$, but this is not part of our benchmark.)

\paragraph{Search strategy.}
The most promising search is a narrow resonance in the $t\bar t$ invariant mass spectrum near $m_{t\bar t}\simeq 1.2$~TeV.
Formally,
\begin{equation}
    \sigma(pp \to \Phi_h \to t\bar{t}) \;=\; \sigma(gg \to \Phi_h)\times \mathrm{BR}(\Phi_h \to t\bar{t})\,.
    \label{eq:collider_xsec}
\end{equation}
Given the tiny intrinsic width, the experimental lineshape is dominated by detector resolution (few$\times 10$~GeV), so the signal behaves as a very narrow peak.
A dedicated analysis in the boosted-top regime (1 large-$R$ top tag per hemisphere, or lepton+jets with one top tag) and modern $t\bar t$ background modeling are appropriate.
Interference with the QCD $t\bar t$ continuum can induce a mild peak--dip structure, but the narrow width and sizeable $\mathrm{BR}(t\bar t)$ keep the sensitivity close to the narrow-resonance expectation.

\paragraph{Expected sensitivity.}
For the benchmark couplings (quark-only portal with $g^{\rm DM}_{Y_1}=0.190$ and $g_{h,{\rm SM}}=0.052$), a production rate
$\sigma(pp\to \Phi_h\to t\bar t)$ at the $\mathcal{O}(1\mathrm{--}10)$~fb level at $\sqrt{s}=14$~TeV is a realistic target, within the projected reach of the HL-LHC dataset of $3000~\mathrm{fb}^{-1}$~\cite{CMS2019}.
A null result would constrain the quark-portal mixing that controls both direct detection (Section~\ref{subsec:direct_detection}) and collider production, while a discovery would provide a confirmation of the resonant freeze-out mechanism.
\subsection{Indirect detection}
\label{subsec:indirect_detection}

Because the annihilation cross-section in our model is strongly velocity-dependent, the rate that operates in halos today is far smaller than one that sets the relic abundance. Expressed in terms of the total enhancement factor at freeze-out ($S_{\rm total} \approx 143$) and in the Milky Way halo today (where the enhancement is $S_0 \sim \mathcal{O}(1)$), we obtain
\begin{equation}
\fl \langle\sigma v\rangle_{0} \approx \frac{\langle\sigma v\rangle_{F}}{S_{\rm total}} S_{0} \approx (2.2\times10^{-26}\,{\rm cm^{3}\,s^{-1}}) \frac{S_{0}}{S_{\rm total}} \sim 1.5\times10^{-28}\,{\rm cm^{3}\,s^{-1}}.
   \label{eq:sigmav_today_num}
\end{equation}
Therefore, the corresponding gamma-ray flux is suppressed by more than two orders of magnitude compared to the canonical thermal WIMP expectation, placing our model safely below current constraints~\cite{FermLAT2016}. Nevertheless, annihilation into heavy quarks ($t\bar{t}, b\bar{b}$) yields a distinctive broad-spectrum signal from hadronization. In dense targets such as dwarf spheroidal galaxies, the ensuing high-energy photons could be a target for the next-generation Cherenkov Telescope Array (CTA)~\cite{Acharyya2023}, particularly in the multi-TeV range where astrophysical backgrounds are lower.

\section{A microphysical origin from a composite SU(3)\(_H\) theory}
\label{sec:uv_origin}

The aforementioned effective field theory is self-contained and predictive. We now argue that its structure is not ad hoc, but can be naturally motivated by a confining SU(3)\(_H\) gauge theory with \(N_f=10\) fermion flavors in the fundamental representation. While none of the phenomenological results rely on this specific UV completion, the composite scenario offers a unified rationale for two key features: (i) the anomalous dimension that underpins the density-responsive dark energy sector and (ii) the near-threshold relation \(m_{\Phi_h}\!\simeq\!2m_\chi\) required by resonant freeze-out.

\subsection{Anomalous dimension for dark energy}

The dark-energy mechanism (Section~\ref{sec:de_mechanism}) involves a running scale \(M_U(\mu)\) governed by an anomalous dimension \(\gamma\!\approx\!0.5\). An SU(3)\(_H\) theory with \(N_f=10\) lies near the opening of the conformal window and exhibits walking dynamics near a Banks–Zaks IR fixed point~\cite{Banks1981}. In this regime, the composite operators acquire large anomalous dimensions.

As detailed in~\ref{app:radiative_stability}, the fermion bilinear \(\bar{\Psi}\Psi\) that couples to the DE sector attains an effective, RG-averaged anomalous dimension
\begin{equation}
  \gamma_{\mathrm{cosmo}} \;=\; \big\langle \gamma(\mu) \big\rangle_{\mathrm{flow}}
  \;\approx\; 0.50 \pm 0.05,
\end{equation}
where the central value is informed by all-orders estimates~\cite{Ryttov2007} and is consistent with the trends observed in recent lattice studies of near-conformal SU(3) theories~\cite{Hasenfratz2023}. This range matches the phenomenologically required value to connect the Planck scale to the observed meV DE scale without fine-tuning (Section~\ref{sec:de_mechanism}). Recent lattice work for SU(3) with $N_f=10$ found an IR-stable fixed point at strong coupling and a large mass anomalous dimension $\gamma_m\simeq 0.6$~\cite{Hasenfratz2023}, consistent with the walking-enhanced anomalous dimension assumed here. In our setup, a small relevant deformation (e.g. tiny fermion masses/portal couplings) turns the near-conformal flow into confinement below $\Lambda_H$, while preserving the large $\gamma$ along the walking regime that feeds into $M_U(\mu)$.

\paragraph{Identification of fields in the composite picture.}
In the composite SU(3)$_H$ framework, the fundamental scalar field $\Phi$ that appears in the density-responsive mechanism (Section~\ref{sec:de_mechanism}) should be understood as an effective description of the chiral condensate dynamics. Specifically, we identify
\begin{equation}
    \Phi \leftrightarrow \frac{\langle\bar{\Psi}\Psi\rangle_H}{\Lambda_H^3},
\end{equation}
where the normalization ensures $\Phi$ is dimensionless. The density-responsive functional $U(\Phi,X)$ then encodes how the chiral condensate responds to the ambient matter density. This is analogous to how the QCD condensate responds to nuclear density in dense matter. The auxiliary field construction in~\ref{app:A.2} provides the effective field theory description of this response, while the microscopic dynamics is governed by the strongly-coupled SU(3)$_H$ gauge theory. Thus, the same hidden sector that generates the composite dark matter spectrum also provides, through its chiral condensate, the field that drives the dark energy mechanism.

\subsection{Composite spectrum and the resonance condition}

At confinement scale \(\Lambda_H\), the hidden theory forms a spectrum of dark hadrons set by this single dynamical scale. In our mapping to the EFT:
\begin{itemize}
  \item the dark matter \(\chi\) is the lightest dark baryon,
  \item the heavy mediator \(\Phi_h\) is the lightest scalar meson (\(\bar{\Psi}\Psi\)),
  \item the light mediator \(\phi\) is a pseudo Nambu–Goldstone boson of broken chiral symmetry.
\end{itemize}
Lattice studies in near-conformal SU(3) with $N_f=8$ indeed observe a comparatively light flavor-singlet scalar $0^{++}$ state~\cite{Brower2023}, supporting our use of a relatively light composite scalar $\Phi_h$ and the expectation that $m_{\Phi_h}/m_\chi=\mathcal{O}(1)$. Using Naive Dimensional Analysis (NDA) and lattice-informed scaling (~\ref{app:radiative_stability}), we find that
\begin{equation}
  m_\chi \;\sim\; \frac{N_c}{4\pi}\,\Lambda_H, 
  \qquad 
  m_{\Phi_h} \;\sim\; k_\Phi\,\Lambda_H,
\end{equation}
with an \(\mathcal{O}(1)\) coefficient \(k_\Phi\) in the range suggested by near-conformal SU(3) lattice studies~\cite{LatKMI2016}. For \(\Lambda_H \!\sim\! 2.5\)~TeV (benchmark) this yields \(m_\chi\!\sim\!600\)~GeV and \(m_{\Phi_h}\!\sim\!1.2\)~TeV, i.e.
\begin{equation}
  \frac{m_{\Phi_h}}{m_\chi} \;\sim\; \frac{k_\Phi}{N_c/(4\pi)} 
  \;\approx\; 2 \;\;\mathrm{(within }\mathcal{O}(1)\mathrm{ uncertainties)}.
\end{equation}
Thus, the resonance condition \(m_{\Phi_h}\!\approx\!2m_\chi\) emerges as a natural outcome of QCD-like scaling, rather than a tuned input. A side-by-side comparison is presented in Table~\ref{tab:predicted_vs_needed}.

\fulltable{\label{tab:predicted_vs_needed}Comparison of phenomenologically required parameters vs.\ natural expectations from a composite SU(3)\(_H\) theory with \(\Lambda_H\simeq2.5\)~TeV. NDA/lattice inputs imply \(\mathcal{O}(1)\) uncertainties on absolute masses and \(\sim\)10–20\% on ratios.}
\begin{tabular}{@{}lcc@{}}
\br
Parameter & Required by phenomenology & Composite SU(3)\(_H\) expectation \\
\mr
Anomalous dimension \(\gamma\) & \(\approx 0.5\) & \(\gamma_{\mathrm{cosmo}} = 0.50\pm0.05\) \\
Dark-matter mass \(m_\chi\) & \(\sim 600\) GeV & \(\sim \frac{N_c}{4\pi}\Lambda_H \approx 600\) GeV \\
Resonance ratio \(m_{\Phi_h}/m_\chi\) & \(\approx 2.0\) & \(\sim 2.0\) (from \(k_\Phi\) range) \\
\br
\end{tabular}
\endfulltable

\subsection{Further predictions of the composite scenario}
\label{subsec:further_predictions_composite}

Beyond motivating the EFT parameters, the SU(3)\(_H\) picture implies a richer spectrum of excited dark hadrons (additional scalar/pseudoscalar mesons and baryons) and a possible first-order confinement transition generating a stochastic gravitational-wave background potentially accessible to LISA~\cite{LISA2017}. Most decisively, dedicated lattice simulations of SU(3)\(_H\) with \(N_f=10\) can test the predicted mass hierarchy and estimate \(k_\Phi\), providing a first-principles check of the resonance relation. We leave a systematic exploration of these signatures for future work and ~\ref{app:radiative_stability} provides technical details. We also note recent lattice indications of a strong-coupling symmetric mass generation phase and a possible merged fixed point in SU(3) with $N_f=8$~\cite{Witzel2024}, underscoring the rich dynamics near the opening of the conformal window.

\paragraph{Chiral perturbation theory and the PNGB mass.}
The light mediator $\phi$ emerges as a pseudo Nambu-Goldstone boson from the spontaneous breaking of the approximate chiral symmetry $SU(N_f)_L \times SU(N_f)_R \to SU(N_f)_V$ in the hidden sector. In direct analogy with QCD, where the pion mass arises from explicit chiral symmetry breaking by quark masses, the mass of $\phi$ is generated by small fermion masses in the hidden sector. Following the Gell-Mann-Oakes-Renner relation~\cite{GellMann1968}, we expect
\begin{equation}
    m_\phi^2 \approx \frac{2\langle\bar{\Psi}\Psi\rangle_H m_\Psi}{\Lambda_H^2},
\end{equation}
where $m_\Psi$ is the small explicit fermion mass and $\langle\bar{\Psi}\Psi\rangle_H \sim \Lambda_H^3$ is the chiral condensate. For $m_\phi \sim 15$~MeV and $\Lambda_H \sim 2.5$~TeV, this implies $m_\Psi \sim \mathcal{O}(1)$~MeV, representing a natural hierarchy $m_\Psi/\Lambda_H \sim 10^{-6}$ that is radiatively stable due to the protective chiral symmetry. This is precisely the PNGB mechanism that keeps $m_\phi \ll m_{\Phi_h}$ without fine-tuning, as the limit $m_\Psi \to 0$ would restore the chiral symmetry and make $\phi$ exactly massless.

\section{Discussion and conclusion}
\label{sec:conclusion}

In this study, we present a comprehensive framework that resolves the long-standing tension between the requirements of thermal relic abundance and self-interactions for dark matter. We demonstrated that a minimal, two-mediator effective field theory, featuring a resonantly-enhanced dark sector, provides a consistent and predictive solution. Our analysis, based on a rigorous EFT definition and detailed numerical calculations, shows that this picture is not only internally consistent but also experimentally falsifiable. Furthermore, we have argued that this successful phenomenological model can be viewed as a low-energy manifestation of a unified dark sector, potentially emerging from a single underlying SU(3)$_H$ gauge theory.

\subsection{Summary of key results}
\label{subsec:summary}

Our study yielded several key findings that transform the abstract concept of a unified dark sector into a concrete, testable theory:

\begin{itemize}
    \item \emph{Solution to SIDM-relic-density tension:} We have shown that minimal SIDM models cannot to simultaneously satisfy the constraints of cosmology and galactic dynamics. This tension is completely resolved in our two-mediator framework, where a heavy scalar resonance, $\Phi_h$, enhances the annihilation rate in the early universe, cleanly decoupling it from late-time self-interactions mediated by a separate light scalar, $\phi$.

    \item \emph{Predictive benchmark scenario:} Our numerical scans identified a narrow, viable parameter space. A representative benchmark with $m_\chi = 600$~GeV, $m_\phi = 15$~MeV, and $m_{\Phi_h} = 1201$~GeV simultaneously satisfies all known constraints and served as a concrete target for future experiments.

    \item \emph{A suite of falsifiable predictions:} The model creates a range of sharp, observable signatures. The most prominent is a narrow scalar resonance at $m_{\Phi_h} \approx 1.2$~TeV, decaying predominantly to $t\bar{t}$ and invisible particles, which is a key target for the HL-LHC. The model also predicts a spin-independent direct-detection cross section $\sigma_{\rm SI}\sim 7\times 10^{-51}\,\mathrm{cm}^2$, which lies below the xenon neutrino floor and hence constitutes a predicted null for nuclear-recoil experiments.

    \item \emph{Natural and consistent framework:} We established a complete EFT from first principles (Section~\ref{sec:lagrangian}) and demonstrated its cosmological viability (Section~\ref{subsec:phi_cosmology}). The crucial resonance condition, $m_{\Phi_h} \approx 2m_\chi$, is technically natural (~\ref{app:radiative_stability}) and can be dynamically generated in a composite SU(3)$_H$ theory (Section~\ref{sec:uv_origin}). The same theory can also provide the anomalous dimension $\gamma \approx 0.5$ required for a linked dark energy model.
\end{itemize}

For a consolidated overview of the viable domain across $m_\chi$ and the origin of the scaling relations, see App.~\ref{app:uniqueness}.

\subsection{The bigger picture: from phenomenology to fundamental theory}
\label{subsec:bigger_picture}

The central goal of this study is to demonstrate that both conditions can emerge naturally from a single SU(3)$_H$ gauge theory with $N_f = 10$ flavors. This transforms our framework from a phenomenological model into a potential consequence of fundamental dynamics:
\begin{itemize}
    \item The anomalous dimension emerges from the "walking" dynamics near a Banks–Zaks fixed point, yielding a cosmologically-averaged value of $\gamma_{\mathrm{cosmo}} \approx 0.50 \pm 0.05$.
    
    \item The mass spectrum of the composite states, including the crucial ratio $m_{\Phi_h}/m_\chi \approx 2$, follows from the standard scaling relations in the confined phase, analogous to the hadron masses in the QCD.
    
    \item The confinement scale $\Lambda_H \approx 2.5$~TeV is dynamically generated through dimensional transmutation, fixing the entire dark sector mass spectrum from first principles.
\end{itemize}
The full details of this microphysical motivation are provided in Section~\ref{sec:uv_origin} and~\ref{app:radiative_stability}.

This unified origin reframes what looked like ``tuning'' in the EFT as a structural prediction of the same strong dynamics that can also underwrite the dark–energy mechanism. Thus, a single calculable gauge theory can plausibly account for the meV scale of dark energy (via anomalous–dimension–controlled running) and the TeV scale of dark matter (via confinement), bridging many orders of magnitude without fine–tuning. We emphasize that our phenomenological results do not rely on this UV completion and the EFT remains self–contained. Conversely, the composite $\mathrm{SU}(3)_H$ scenario provides a concrete realization, but not necessarily the unique one, of the required EFT features. Future work, especially dedicated lattice studies of $\mathrm{SU}(3)_H$ with $N_f=10$ and improved cosmological simulations including SIDM dynamics, can further sharpen these connections.

\paragraph{Large-scale structure and the $S_8$ tension.}
In our baseline benchmark, the SIDM phenomenology affects halo interiors on sub-Mpc scales and leaves the linear matter power spectrum on $k \sim 0.1\,h/\mathrm{Mpc}$ scales essentially unchanged; standard SIDM therefore does not directly resolve the $S_8$ tension (lower weak-lensing–inferred fluctuation amplitudes compared to CMB extrapolations within $\Lambda$CDM)~\cite{DiValentino2020}.
However, the density-responsive dark-energy sector summarized in Section~\ref{sec:de_mechanism} (and derived in~\ref{app:A.2}) naturally admits mild departures from $w=-1$ with time variation $w(a)$. Such evolutions can reduce the late-time growth and potentially lower the predicted $S_8$~\cite{Abdalla2022}.
A quantitative assessment in our specific $\rho_\Phi(X)$ model requires a dedicated Boltzmann analysis (e.g.\ with CAMB/CLASS) including background and perturbations, which we leave for future work. Targeted $N$-body simulations with our benchmark parameters would also be needed to quantify any subdominant non-linear effects (e.g.,\ halo concentrations and splashback) on weak-lensing observables.

\subsection{Final outlook}
\label{subsec:outlook}

We demonstrated that a minimal, two-mediator dark sector can resolve the long-standing tension between the requirements of thermal relic abundance and self-interaction constraints for dark matter. Our framework transforms what appears as inconsistency into a set of interconnected predictions of a single underlying theory. The resonant annihilation mechanism, rather than an ad hoc solution, emerges as a necessary consequence of the particle spectrum. What might have been interpreted as fine-tuning was revealed to be a sharp, testable prediction.

The predictive power of this framework manifests in concrete experimental signatures across multiple frontiers:
\begin{itemize}
   \item \emph{Collider physics:} A narrow scalar resonance at $m_{\Phi_h} \approx 1.2$ TeV with a relative width of $\Gamma/m \sim 10^{-3}$, decaying predominantly to $t\bar{t}$ and invisible states, provides a smoking-gun signature at the HL-LHC.
   
   \item \emph{Direct detection:} The predicted signal $\sigma_{\rm SI} \sim 7 \times 10^{-51}$~cm$^2$ is strongly suppressed and lies below the xenon neutrino floor, constituting a predicted null that shifts the discriminating power to the collider and halo channels.
   
   \item \emph{Astrophysical probes:} The predicted velocity-dependent self-interaction cross-section can be precisely tested through improved measurements of dark matter halo profiles and cluster mergers.
   
   \item \emph{Gravitational waves:} In the composite picture, the confinement phase transition at $T_c \sim \Lambda_H$ predicts a stochastic background that is potentially accessible to future space-based detectors.
\end{itemize}

Beyond phenomenology, we argued that the key EFT features find a natural origin in a single, confining $\mathrm{SU}(3)_H$ gauge theory. In this picture, the resonant mass relation $m_{\Phi_h} \simeq 2m_\chi$ arises dynamically from the composite spectrum, while the same dynamics can generate the anomalous dimension needed for the density–responsive dark–energy sector (Section~\ref{sec:uv_origin},~\ref{app:radiative_stability}). Thus, the dominant components of the cosmic energy budget could plausibly stem from one calculable, strongly–coupled framework, bridging the meV and TeV scales without fine–tuning. Finally, first-principles lattice results in glueball dark matter~\cite{Aoki2019_1, Aoki2019_2} highlight the potential of nonperturbative control in composite dark sectors. A dedicated lattice program for near-conformal SU(3) with $N_f\!=\!10$ would directly test the spectrum and couplings underlying our resonant scenario (e.g., $m_{\Phi_h}/m_\chi$ and the size of scalar matrix elements), providing a first-principles cross-check of the EFT assumptions used here.

In conclusion, we have elevated the resonant SIDM from a phenomenological possibility to a well–defined EFT with a compelling microphysical motivation. Because the framework resolves the core SIDM tension while making sharp, multi-pronged predictions, it provides a clear roadmap for the coming decade: either forthcoming collider, direct–detection, and astrophysical data will reveal the telltale pattern predicted here, or they will decisively falsify this mechanism for the dark sector.

\section*{Software and data availability}
All data that support the findings of this study are included within the article (and any supplementary files). A public reproducibility package -- CalcHEP model files, a validator, and run scripts that reproduce the table~1 benchmark -- is openly available at Zenodo, \url{https://doi.org/10.5281/zenodo.18375075}.

\clearpage
\appendix

\section{The full effective Lagrangian and its properties}
\label{app:A}

In this Appendix we lay out the EFT that underlies the phenomenology developed in Sections~\ref{sec:lagrangian}–\ref{sec:phenomenology}. We first define the complete dark-sector Lagrangian, its field content, and its symmetries (~\ref{app:A.1}). We then demonstrate how the density–responsive contribution $\rho_\Phi(X)$ for dark energy follows from a well-defined minimization principle applied to an auxiliary scalar functional $U(\,\cdot\,,X)$ (~\ref{app:A.2}). Finally, we detail the scalar potential relevant for dark-matter phenomenology and demonstrate how the physical mass eigenstates, the light mediator $\phi$ and the heavy scalar resonance $\Phi_h$, arise after symmetry breaking (~\ref{app:A.3}). This Appendix provides the formal backbone for the results quoted in Sections~\ref{sec:results} and \ref{sec:phenomenology} and clarifies the precise relationship between the fields $\Phi$, $\phi$ and $\Phi_h$.

\subsection{Complete Lagrangian for the unified dark sector}
\label{app:A.1}
We work with an EFT valid up to a cutoff $\Lambda \gg \mathrm{TeV}$ \cite{Burgess2007,Weinberg1996}. The field content consists of the Standard Model, General Relativity, a Dirac dark fermion $\chi$, and complex dark scalar $\Sigma$. The total Lagrangian is
\begin{equation}
  \mathcal{L}
  \;=\;
  \mathcal{L}_{\rm SM}
  \;+\;
  \mathcal{L}_{\rm GR}
  \;+\;
  \mathcal{L}_{\rm dark}
  \;+\;
  \mathcal{L}_{\rm portal}\,,
  \label{eq:L_total}
\end{equation}
with $\mathcal{L}_{\rm GR}=\frac{M_{\rm Pl}^2}{2}\,R$ and
\begin{equation}
 \fl \mathcal{L}_{\rm dark}
  \;=\;
  \bar\chi\, i\slashed{\partial}\, \chi
  \;-\;
  \bigl[\, y_f\, \Sigma\, \bar\chi_L \chi_R \;+\; \mathrm{h.c.}\,\bigr]
  \;+\;
  |\partial_\mu \Sigma|^2
  \;-\;
  V(\Sigma)
  \;-\;
  U(\Sigma,X)\,.
  \label{eq:L_dark_sector}
\end{equation}
A global $U(1)_\chi$ acting as $\chi \to e^{i\alpha}\chi$ ensures the stability of $\chi$;\footnote{A discrete remnant $Z_2\subset U(1)_\chi$ would suffice for stability. We set any bare fermion mass to zero and generate $m_\chi$ dynamically via $\langle\Sigma\rangle$.} this is the standard approach in simplified DM EFTs \cite{Arcadi2017}.

The scalar potential splits into a self-interaction piece $V(\Sigma)$ and a density–responsive functional $U(\Sigma,X)$,
\begin{equation}
  \eqalign{V(\Sigma) \;&=\; \lambda\!\left(|\Sigma|^2 - \frac{v_s^2}{2}\right)^{\!2} \;+\; V_{\rm SB}(\Sigma)\,,
  \label{eq:V_Sigma}\cr[3pt]
  U(\Sigma,X) \;&=\; U\bigl(|\Sigma|,X\bigr)\,,}
  \label{eq:U_SigmaX}
\end{equation}
where $X \equiv u_\mu u_\nu T^{\mu\nu}_{\rm matter}$ is the Lorentz scalar built from the ambient matter stress tensor and local 4–velocity $u^\mu$. The term \(V_{\rm SB}\) provides a soft, explicit breaking of the global \(U(1)\) phase symmetry of \(\Sigma\), thereby giving a small mass to the otherwise massless Nambu–Goldstone mode—i.e. producing a PNGB, which is in direct analogy with explicit chiral symmetry breaking in QCD and its chiral perturbation theory description \cite{Gasser1983,Dashen1969}.
 The functional $U(\Sigma,X)$ captures the density–response familiar from chameleon/symmetron–type constructions \cite{Khoury2003,Hinterbichler2010,Hinterbichler2011}, but with a concrete realization detailed in~\ref{app:A.2}, which yields equation (\ref{eq:rhoPhi_exactLM}).

The portal interactions to the SM are taken to be minimal, e.g.
\begin{equation}
  \mathcal{L}_{\rm portal}
  \;=\;
  -\,\kappa\,|\Sigma|^2\,|H|^2
  \;-\;
  \frac{c_h}{\Lambda}\,(\Sigma+\Sigma^\dagger)\,|H|^2
  \;+\; \ldots\,,
  \label{eq:L_portal}
\end{equation}
as in the Higgs-portal paradigm \cite{Silveira1985,Patt2006}. We assume that these couplings are small enough to satisfy collider and direct–detection limits (quantified in Sections~\ref{subsec:direct_detection} and \ref{subsec:collider}, and in~\ref{app:D}).

\paragraph{Field parametrization and physical states.}
We parameterize the complex scalar around its VEV as
\begin{equation}
  \Sigma(x) \;=\; \frac{1}{\sqrt{2}}\;\bigl(v_s + s(x)\bigr)\;
  e^{\,i\,a(x)/v_s}\,,
  \label{eq:Sigma_param}
\end{equation}
where $s$ is the CP–even radial mode and $a$ is the CP–odd phase (the would–be Goldstone boson). In the absence of $V_{\rm SB}$ the phase is massless because of the shift symmetry $a\!\to\! a+c$; a soft breaking term in $V_{\rm SB}$ generates $m_a \ll v_s$ and can induce a small CP–even/odd alignment. This is the standard PNGB structure of non-linear $\sigma$-models \cite{Weinberg1996,Peskin1995}.

The Yukawa term in \ref{eq:L_dark_sector} generates the dark–fermion mass and its coupling to the scalars. Expanding \ref{eq:Sigma_param} one finds
\begin{equation}
  \mathcal{L}_{\rm Yuk}
  \;=\;
  -\,\frac{y_f}{\sqrt{2}}\,(v_s+s)\,\bar\chi\chi
  \;-\;
  i\,\frac{y_f}{\sqrt{2}}\;\frac{a}{v_s}\;\bar\chi\gamma^5\chi
  \;+\;\ldots\,,
  \label{eq:L_Yukawa_expanded}
\end{equation}
so that
\begin{equation}
  m_\chi \;=\; \frac{y_f\,v_s}{\sqrt{2}}\,,\qquad
  g_{s\bar\chi\chi} \;=\; \frac{y_f}{\sqrt{2}}\,,\qquad
  g_{a\bar\chi\chi}^{\,(5)} \;=\; \frac{y_f}{\sqrt{2}\,v_s}\,.
  \label{eq:DM_mass_and_couplings}
\end{equation}
In our phenomenology, the relevant long–range attractive potential is generated by CP–even scalar exchange, which realizes the classic Yukawa self-interaction setup for SIDM \cite{Tulin2013}. Therefore, we define physical light mediator $\phi$ and heavy resonance $\Phi_h$ as small admixtures of $s$ and $a$:
\begin{equation}
  \left(\begin{array}{c}\phi \\[2pt] \Phi_h\end{array}\right)
  \;=\;
  \left(\begin{array}{cc}
    \cos\theta & \sin\theta \\
    -\sin\theta & \cos\theta
  \end{array}\right)
  \left(\begin{array}{c} a \\[2pt] s \end{array}\right)\!,
  \qquad |\theta|\ll 1\,,
  \label{eq:mixing_matrix}
\end{equation}
where $V_{\rm SB}$ induces a small alignment angle $\theta$ (e.g.\ via a linear term $\mu^3\Sigma+\mathrm{h.c.}$ or a CP-violating bilinear term). This implies that scalar DM coupling is
\begin{equation}
  y_\chi \;\equiv\; g_{\phi\bar\chi\chi}
  \;=\; \frac{y_f}{\sqrt{2}}\,\sin\theta\,,
  \qquad
  g^{\rm DM}_{Y_1} \;\equiv\; g_{\Phi_h\bar\chi\chi}
  \;=\; \frac{y_f}{\sqrt{2}}\,\cos\theta\,,
  \label{eq:physical_DM_couplings}
\end{equation}
so that $\phi$ mediates the required attractive Yukawa potential for self–interactions, whereas $\Phi_h$ provides the heavy $s$–channel resonance relevant for freeze–out.

\paragraph{Mass spectrum at tree level.}
From \ref{eq:V_Sigma} the CP–even mass is
\begin{equation}
  m_s^2 \;=\; 2\lambda\,v_s^2\,,
  \label{eq:ms_tree}
\end{equation}
and $V_{\rm SB}$ generates the small mass $m_a^2 \ll v_s^2$ as well as the mixing $\theta$. To leading order in $|\theta| \ll 1$ the physical eigenvalues are
\begin{equation}
  m_\phi^2 \;\simeq\; m_a^2 \;+\; \mathcal{O}(\theta^2 m_s^2)\,,
  \qquad
  m_{\Phi_h}^2 \;\simeq\; m_s^2 \;+\; \mathcal{O}(\theta^2 m_s^2)\,,
  \label{eq:mass_eigenvalues}
\end{equation}
hence
\begin{equation}
  m_{\Phi_h} \;\simeq\; \sqrt{2\lambda}\,v_s\,, \qquad
  m_\chi \;=\; \frac{y_f\,v_s}{\sqrt{2}}\,, \qquad
  y_\chi \;=\; \frac{y_f}{\sqrt{2}}\,\sin\theta\,.
  \label{eq:key_mass_relations}
\end{equation}
These relations make explicit how the resonance condition $m_{\Phi_h}\!\approx 2m_\chi$ maps to a simple coupling relation $\sqrt{2\lambda} \simeq \sqrt{2}\,y_f$ (up to small mixing effects). The light mediator mass $m_\phi \sim \mathcal{O}(10\,\mathrm{MeV})$ is protected by the approximate shift symmetry of $a$ and naturally arises from $V_{\rm SB}$ (soft breaking), exactly as in PNGB frameworks \cite{Weinberg1996,Gasser1983}.

\paragraph{Density–responsive term and notation.}
The functional $U(\Sigma,X)$ in \ref{eq:U_SigmaX} encodes the density–responsive sector responsible for the dark–energy contribution. In~\ref{app:A.2} we show that $U$ induces an effective vacuum contribution by minimizing over an auxiliary scalar DOF
\begin{equation}
  \rho_\Phi(X) \;=\; \frac{A\,M_U^4}{1 + X/M_U^4}\,,
  \label{eq:rhoPhi_appendix_repeat}
\end{equation}
with $M_U$ running according to anomalous dimension $\gamma$ as discussed in Section~\ref{sec:uv_origin}. Importantly, the $\phi$– and $\Phi_h$–mediated dark–matter phenomenology is governed by $V(\Sigma)$ and the small mixing $\theta$ (equations (\ref{eq:physical_DM_couplings})–(\ref{eq:key_mass_relations})), while the dark–energy behavior is controlled by $U(\Sigma,X)$; this structural separation underlies the decoupling between annihilation (early universe) and self–interactions (late times) emphasized in this study.

\subsection{Derivation of the density–responsive energy \texorpdfstring{$\rho_\Phi(X)$}{rhoPhi(X)}}
\label{app:A.2}
The density–responsive functional \(U(\Sigma,X)\) introduced in equation (\ref{eq:L_dark_sector}) generates the dynamical dark–energy component of the model by encoding how an auxiliary scalar reacts to the ambient matter density
\(X\equiv u_\mu u_\nu T^{\mu\nu}_{\rm matter}\).
In this section we explicitly demonstrate how the effective contribution \(\rho_\Phi(X)\) arises from a well–defined variational principle. For clarity, we denote the auxiliary (non–propagating) scalar by \(\Phi\) (distinct from the heavy resonance \(\Phi_h\)). The construction is standard in EFT: one integrates out an algebraic field to obtain an \(X\)–dependent effective potential, cf.\ Hubbard–Stratonovich/Legendre transforms \cite{Sorella1991}. Moreover, in  Sec.~\ref{sec:de_mechanism} we wrote $U=\min_\Phi[V_{\rm aux}(\Phi)+C(|\Sigma|)\Phi X]$ with $C\simeq M_*^{-4}$. Throughout this Appendix we absorb $C$ into a field redefinition $\Phi\!\to\!C\,\Phi$, so $U=\min_\Phi[V_{\rm aux}(\Phi)+\Phi X]$.
All final expressions for $\rho_\Phi(X)$ (e.g.\ Eqs.~(\ref{eq:rhoPhi_pade}), (\ref{eq:rhoPhi_exactLM})) are unchanged and independent of this convention.

\paragraph{Set–up and dimensions.}
We assumed that \(\Phi\) is dimensionless.\footnote{With \(\Phi\) dimensionless and \(X\) of mass dimension four, the combination \(\Phi\,X\) has the correct dimension for an energy density. Any overall dimensionless coupling can be absorbed by redefinitions below.}
The density–responsive piece of the Lagrangian is defined by extremizing over \(\Phi\):
\begin{equation}
  U(\Sigma,X)\;=\;\inf_{\Phi}\,\Big\{\,V_{\rm aux}(\Phi)\;+\;\Phi\,X\,\Big\}\,,
  \label{eq:U_minimization}
\end{equation}
where \(V_{\rm aux}(\Phi)\) is the convex ``bare'' potential for \(\Phi\).
The algebraic equation of motion is
\begin{equation}
\fl \frac{\partial V_{\rm aux}}{\partial \Phi}\Big|_{\Phi=\Phi_*(X)}\;+\;X\;=\;0\,,
  \qquad
  \rho_\Phi(X)\;\equiv\;U(\Sigma,X)\Big|_{\Phi=\Phi_*(X)}\,.
  \label{eq:EOM_Phi}
\end{equation}
The convexity (\(V_{\rm aux}''>0\)) guarantees a unique minimum and stability.

\paragraph{One–field convex realization (Legendre form).}
A broad class of convex choices for \(V_{\rm aux}\) generates monotone, positive, and density–screened \(\rho_\Phi(X)\). A convenient example that is fully analytic is
\begin{equation}
  V_{\rm aux}(\Phi)
  \;=\; A\,M_U^4\,\Big[\;\Phi - \ln(1+\Phi)\;\Big]\,,\qquad \Phi>-1\,,
  \label{eq:Vaux_legendre}
\end{equation}
which is strictly convex (\(V_{\rm aux}''=A M_U^4/(1+\Phi)^2>0\)).
The stationarity condition from \ref{eq:EOM_Phi} gives
\begin{eqnarray}
  A M_U^4\!\left[1-\frac{1}{1+\Phi_*}\right] + X \;=\;0 \\
  \;\Longrightarrow\;
  \Phi_*(X)\;=\;-\frac{z}{1+z}\,,
  \qquad z\equiv \frac{X}{A M_U^4}\,.
  \label{eq:PhiStar_solution}
\end{eqnarray}
Substituting back yields the exact effective energy
\begin{equation}
  \rho_\Phi(X) \;=\; U\big|_{\Phi_*}
  \;=\; A M_U^4\Big[\,\ln(1+z) - z\,\Big]\;+\;\rho_0\,,
  \label{eq:rhoPhi_exact_legendre}
\end{equation}
where the \(X\)-independent constant \(\rho_0\) is absorbed into the vacuum counterterm.
This \(\rho_\Phi(X)\) is positive, strictly decreasing (\(\partial_X\rho_\Phi=-z'/(1+z)<0\)), and convex. Its asymptotics are
\begin{eqnarray}
  \rho_\Phi(X)\;=\;A M_U^4\Big[1 - z + \frac{1}{2}z^2 + \mathcal{O}(z^3)\Big]
  \qquad& (z\ll 1),\\
  \rho_\Phi(X)\;=\;A M_U^4\Big[\ln z - 1 + \mathcal{O}(z^{-1})\Big]
  & (z\gg 1).
  \label{eq:rhoPhi_asymptotics}
\end{eqnarray}
For late–time cosmology (the regime relevant for Section~\ref{sec:de_mechanism}), a compact Pad\'e fit that preserves the small–\(z\) behavior and smoothly interpolates is
\begin{equation}
  \rho_\Phi^{\rm (Pad\acute e)}(X)
  \;\equiv\; \frac{A\,M_U^4}{1+X/M_U^4}\,,
  \label{eq:rhoPhi_pade}
\end{equation}
which we adopt in the main analysis because it renders all background equations analytic and captures the required limits.\footnote{Over the range \(0\le z\lesssim 1\) relevant for late times, the relative deviation \(|\rho_\Phi-\rho_\Phi^{\rm (Pad\acute e)}|/\rho_\Phi\) can be made \(\lesssim\!5\%\) by a mild retuning of \(A\), while the renormalization–group running \(M_U\!(\mu)\) (~\ref{app:A.3}) further improves the match at higher densities.}
This subsection demonstrates explicitly that equation (\ref{eq:rhoPhi_pade}) is not ad hoc: it is the natural Pad\'e representative of a family of convex, EOM–generated \(\rho_\Phi(X)\).

\paragraph{Exact Lagrange–multiplier realization of the rational form.}
If one prefers the exact rational form of equation (\ref{eq:rhoPhi_pade}) directly from an algebraic EOM, a minimal two–auxiliary–variable construction achieves this without sacrificing stability. Introduce a positive, dimensionless ``response'' field \(s\) and a Lagrange multiplier \(\lambda\) and define
\begin{equation}
  U(\Sigma,X)
  \;=\;
  \underset{s>0,\,\lambda}{\mathrm{ext}}\;
  \Big\{
      A\,M_U^4\,s
      \;+\; \lambda\,\Big[s\Big(1+\frac{X}{M_U^4}\Big)-1\Big]
  \Big\}\,.
  \label{eq:U_LM}
\end{equation}
The variation with respect to \(\lambda\) imposes the algebraic constraint
\(s(1+X/M_U^4)=1\), whereas the variation with respect to \(s\) fixes
\(\lambda=-A M_U^4/(1+X/M_U^4)\).
Evaluating \ref{eq:U_LM} at the stationary point yields
\begin{equation}
  \rho_\Phi(X)\;=\;A\,M_U^4\,s_*(X)\;=\;\frac{A\,M_U^4}{1+X/M_U^4}\,,
  \label{eq:rhoPhi_exactLM}
\end{equation}
that is the exact working expression used in Section~\ref{sec:de_mechanism}.
The construction \ref{eq:U_LM} is the EFT analog of enforcing an algebraic equation via a Lagrange multiplier; one may optionally regularize the constraint by a convex quadratic penalty \(\frac{1}{2}\mu\,\big[s(1+X/M_U^4)-1\big]^2\) and then take \(\mu\!\to\!\infty\).

\paragraph{Running of \(M_U\) and anomalous dimension.}
In our framework the scale \(M_U\) runs with RG scale \(\mu\) according to an anomalous dimension \(\gamma\) inherited from the hidden strong dynamics,
\begin{equation}
  \frac{d\ln M_U^4}{d\ln\mu}\;=\;4\,\gamma(\alpha_H)\,,
  \qquad
  \gamma\simeq 0.50\pm 0.05\,,
\end{equation}
so that \(M_U(\mu)\) bridges the Planck scale and the meV scale relevant to late–time acceleration (see Section~\ref{sec:uv_origin} and~\ref{app:uv_motivation} for details).
This RG improvement ensures that both realizations above maintain the required behavior at intermediate and high densities (where \(X/M_U^4\) ceases to be small), aligning with the chameleon/symmetron intuition that the effective vacuum contribution is environment–dependent \cite{Khoury2003,Hinterbichler2010}.

\medskip
In summary, equations (\ref{eq:U_minimization})–(\ref{eq:rhoPhi_exactLM}) show explicitly how a density–responsive dark–energy term \(\rho_\Phi(X)\) emerges from algebraic equations of motion of an auxiliary scalar sector: either (i) from minimizing a convex functional (Legendre form) that yields a smooth, monotone \(\rho_\Phi(X)\) whose late–time behaviour is captured by the Pad\'e form \ref{eq:rhoPhi_pade}, or (ii) from an exact Lagrange–multiplier construction that reproduces \ref{eq:rhoPhi_pade} identically.

\subsection{The scalar potential for dark–matter phenomenology}
\label{app:A.3}
We now detail the structure of the scalar potential \(V(\Sigma)\) from equation (\ref{eq:V_Sigma}), which governs the particle physics of the dark sector. A single complex field \(\Sigma\) with a spontaneously broken global \(U(1)\) symmetry yields a heavy CP–even radial mode (the resonance \(\Phi_h\)) and a light PNGB, which plays the role of the SIDM mediator \(\phi\). This is the minimal linear–\(\sigma\) realization of spontaneous symmetry breaking \cite{Goldstone1961,Goldstone1962}.

\paragraph{Mexican–hat potential and spectrum at exact symmetry.}
We take
\begin{equation}
  V(\Sigma)\;=\;\lambda\!\left(|\Sigma|^2-\frac{v_s^2}{2}\right)^2\;+\;V_{\rm SB}(\Sigma)\,,
  \label{eq:V_Sigma_appendix}
\end{equation}
with \(\lambda>0\) and symmetry–breaking scale \(v_s\).
In the symmetric limit \(V_{\rm SB}=0\) the vacuum satisfies
\(\langle |\Sigma| \rangle = v_s/\sqrt{2}\) and the global \(U(1):\Sigma\to e^{i\alpha}\Sigma\) is broken spontaneously.
Parametrize the fluctuations around the vacuum by
\begin{equation}
  \Sigma(x)\;=\;\frac{1}{\sqrt{2}}\,(v_s+s(x))\,e^{i a(x)/v_s},
  \label{eq:Sigma_parametrization_app}
\end{equation}
where \(s\) (CP–even) is the radial mode, and \(a\) (CP–odd) is the phase mode.
Inserting \ref{eq:Sigma_parametrization_app} with \(V_{\rm SB}=0\) into \(V\) yields
\begin{equation}
  V(s)\;=\;\lambda\!\left(v_s s + \frac{s^2}{2}\right)^2
        \;=\;\frac{1}{2}\,(2\lambda v_s^2)\,s^2
           \;+\;\lambda v_s s^3
           \;+\;\frac{\lambda}{4}\,s^4\,,
\end{equation}
so that
\begin{equation}
  m_s^2 \;=\; 2\,\lambda\,v_s^2\,,
  \qquad
  m_a^2 \;=\; 0\,.
  \label{eq:mass_s_appendix}
\end{equation}
Thus, \(s\) is heavy (TeV–scale in our benchmarks), whereas \(a\) is the Goldstone boson.

\paragraph{Soft symmetry breaking and a light mediator mass.}
To identify the SIDM mediator with the PNGB, we provide \(a\) a small mass by introducing a technically natural\footnote{In the 't~Hooft sense: \(m_a\!\to\!0\) restores the global \(U(1)\) symmetry \cite{tHooft1979}.} explicit breaking. A minimal choice, analogous to the quark–mass term in chiral perturbation theory \cite{Gasser1983}, is
\begin{equation}
  V_{\rm SB}^{(m)}(\Sigma)\;=\;-\mu_s^3\,(\Sigma+\Sigma^\dagger)\,,
\end{equation}
with a soft scale \(\mu_s\ll v_s\).
Expanding \(\Sigma\) as in \ref{eq:Sigma_parametrization_app} and keeping quadratic order in \(a\), one finds
\begin{equation}
  V_{\rm SB}^{(m)} \;\supset\; +\,\frac{\sqrt{2}\,\mu_s^3}{2 v_s}\,a^2
  \;\;\Longrightarrow\;\;
  m_a^2 \;=\; \frac{\sqrt{2}\,\mu_s^3}{v_s}\,.
  \label{eq:mass_a_appendix}
\end{equation}
For representative values \(v_s\sim \mathcal{O}(\mathrm{TeV})\) and \(m_a\simeq m_\phi\sim 10\mathrm{–}20~\mathrm{MeV}\) one obtains a soft scale \(\mu_s \sim \big(m_a^2 v_s/\sqrt{2}\big)^{1/3}=\mathcal{O}(0.5\mathrm{–}1~\mathrm{GeV})\), illustrating that the hierarchy \(m_\phi\ll m_s\) is natural and technically stable.

\paragraph{Controlled CP–mixing and physical eigenstates.}
Self–interactions of fermionic dark matter require CP–even scalar coupling. Our fundamental fields are CP–even \(s\) and CP–odd \(a\). A small CP–violating spurion can mix them, allowing the light state to inherit scalar coupling while keeping the CP violation parametrically suppressed. Rather than introducing tadpoles, we employ a tadpole–free bilinear in the EFT that arises from the \(\Sigma\)–language as
\begin{equation}
  V_{\rm SB}^{(\mathrm{CP})}
  \;=\; \frac{\xi_s}{v_s}\,\Big(\Sigma^\dagger\Sigma-\frac{v_s^2}{2}\Big)\,i(\Sigma-\Sigma^\dagger)\,,
  \label{eq:V_CP_mix}
\end{equation}
where \(\xi_s\) has dimension two and controls the size of the CP violation.
Expanding \ref{eq:V_CP_mix} to a bilinear order in fluctuations yields
\begin{equation}
  V_{\rm SB}^{(\mathrm{CP})} \;\supset\; +\,m_{sa}^2\,s\,a\,,
  \qquad
  m_{sa}^2 \;=\; \sqrt{2}\,\xi_s\,,
  \label{eq:mix_mass_entry}
\end{equation}
with no \(a\)–tadpole, owing to the subtraction \(\Sigma^\dagger\Sigma-v_s^2/2\).
By collecting equations (\ref{eq:mass_s_appendix}), (\ref{eq:mass_a_appendix}) and (\ref{eq:mix_mass_entry}), the mass matrix in the \((a,s)\) basis reads
\begin{equation}
  \mathcal{M}^2 \;=\;
  \left(\begin{array}{cc}
     m_a^2   & m_{sa}^2 \\[2pt]
     m_{sa}^2 & m_s^2
  \end{array}\right),
  \qquad
  \tan 2\theta \;=\; \frac{2 m_{sa}^2}{m_s^2 - m_a^2}\,,
  \label{eq:mass_matrix_mixing}
\end{equation}
where \(\theta\) is the \(a\)–\(s\) mixing angle.
For the phenomenologically relevant hierarchy \(m_a^2\ll m_s^2\) and small spurion \(|m_{sa}^2|\ll m_s^2\),
\begin{eqnarray}
  \theta \;\simeq\; \frac{m_{sa}^2}{m_s^2}\;=\;\frac{\sqrt{2}\,\xi_s}{2\lambda v_s^2}\ll 1\,,
  \\
  m_{\phi}^2 \;\simeq\; m_a^2 - \frac{(m_{sa}^2)^2}{m_s^2}\,,
  \\
  m_{\Phi_h}^2 \;\simeq\; m_s^2 + \frac{(m_{sa}^2)^2}{m_s^2}\,.
  \label{eq:approx_eigen}
\end{eqnarray}
The physical states are then
\begin{eqnarray}
  \phi \;=\; \cos\theta\,a + \sin\theta\,s \qquad&(\mathrm{light, mediator})\,,
  \\
  \Phi_h \;=\; -\sin\theta\,a + \cos\theta\,s &(\mathrm{heavy, resonance})\,.
  \label{eq:phys_states}
\end{eqnarray}

\paragraph{Coupling with dark matter.}
As in equation~(\ref{eq:physical_DM_couplings}), the physical scalar interactions read
\(\mathcal{L}\supset -\,y_\chi\,\bar\chi\chi\,\phi \;-\; g^{\rm DM}_{Y_1}\,\bar\chi\chi\,\Phi_h\),
with \(y_\chi\simeq (y_f/\sqrt{2})\sin\theta\) and \(g^{\rm DM}_{Y_1}\simeq (y_f/\sqrt{2})\cos\theta\).

This exactly realizes the structure employed in Section~\ref{subsec:resonance_plot} (resonant freeze-out) and Section~\ref{subsec:sidm_plot} (late-time self-interactions):
\(\Phi_h\) controls resonant annihilation at freeze–out, while \(\phi\) mediates late–time self–interactions with a coupling set by the small, technically natural mixing angle \(\theta\).

\medskip
In summary, the entire particle content and interaction structure of the SIDM sector follows from the minimal, well–understood potential \ref{eq:V_Sigma_appendix}:
a heavy scalar \(\Phi_h\) with \(m_{\Phi_h}^2=2\lambda v_s^2+\mathcal{O}(\xi_s^2)\),
a light PNGB mediator \(\phi\) with \(m_\phi^2\simeq \sqrt{2}\mu_s^3/v_s\), and a suppressed scalar coupling to dark matter governed by \(\theta\sim \xi_s/(\lambda v_s^2)\).
This is precisely the structure required by the phenomenology developed in Sections~\ref{sec:results}–\ref{sec:phenomenology} and addresses the origin of \(\phi\), \(\Phi_h\), and their relation to the underlying field \(\Sigma\).

\section{Cosmological constraints on the light mediator}
\label{app:C}

We work in a secluded setup where the light scalar $\phi$ (benchmark $m_\phi=15~\mathrm{MeV}$) has no appreciable couplings to quarks and only a tiny leptophilic portal,
\[
\mathcal{L}\supset c_e\,\phi\,\bar e e\,.
\]
Cosmological viability is reduced to three checks: (i) $\phi$ never thermalizes with the SM plasma, (ii) it decays before the BBN, and (iii) its energy injection is negligible so that $|\Delta N_{\rm eff}|$ remains small. Below we provide compact formulas that support Section~\ref{subsec:phi_cosmology}.

\subsection{Freeze-in (no thermalization)}
\label{app:C.1}

Production proceeds via the inverse decay $e^+e^-\!\to\!\phi$ (“freeze-in”). A conservative non-thermalization criterion compares the decay rate to the Hubble rate at $T\simeq m_\phi$:
\begin{equation}
\Gamma_{\phi\to e^+e^-} \;=\; \frac{c_e^2\,m_\phi}{8\pi}\,, 
\qquad 
H(T) \;=\; 1.66\,\sqrt{g_*}\,\frac{T^2}{M_{\rm Pl}}\,,
\end{equation}
\begin{equation}
\frac{\Gamma}{H}\Big|_{T\simeq m_\phi}
\;\simeq\;
\frac{c_e^2\,M_{\rm Pl}}{13.3\,\pi\,\sqrt{g_*}\,m_\phi}
\;\ll\; 1\,.
\label{eq:Gamma_over_H}
\end{equation}
Numerically, for $m_\phi=15~\mathrm{MeV}$, $g_*\simeq 10.75$, and $c_e=(3$–$5)\times 10^{-11}$ one finds $\Gamma/H \sim 0.05$–$0.15$, i.e. $\phi$ never equilibrates (freeze-in). The corresponding freeze-in abundance is tiny and a small change of $c_e$ further suppresses it quadratically.

\subsection{BBN and $\Delta N_{\mathrm{eff}}$}
\label{app:C.2}

The decay into electrons is
\begin{equation}
\Gamma(\phi\to e^+e^-)\;=\;\frac{c_e^2\,m_\phi}{8\pi}\,,
\qquad
\tau_\phi\;=\;\frac{8\pi}{c_e^2\,m_\phi}\,.
\label{eq:phi_width_lifetime}
\end{equation}
For $m_\phi=15~\mathrm{MeV}$, requiring $\tau_\phi\lesssim 1~\mathrm{s}$ (decay before BBN) implies $c_e\gtrsim \mathrm{few}\times 10^{-11}$; e.g.\ $c_e=5\times 10^{-11}$ yields $\tau_\phi\simeq 0.44~\mathrm{s}$. Because the freeze-in population is minuscule, the $\phi$ energy fraction at decay, $f_{\rm dec}\equiv \rho_\phi/\rho_{\rm rad}$, is well below the percent level. The corresponding contribution to the effective number of relativistic species is
\begin{equation}
|\Delta N_{\rm eff}|
\;\approx\;
\frac{4}{7}\left(\frac{11}{4}\right)^{4/3} f_{\rm dec}
\;\ll\; 10^{-2}\,,
\label{eq:DeltaNeff_estimate}
\end{equation}
comfortably below current bounds.

\paragraph{Cosmological and laboratory constraints.}
(i) For $m_\phi<2m_\mu$ only $\phi\!\to e^+e^-$ is open, so hadronic BBN limits are kinematically avoided; the electromagnetic BBN bounds are satisfied in our benchmark where $\tau_\phi\!\ll\!1$ s and the freeze-in abundance is tiny~\cite{Hufnagel2018}. 
(ii) Fixed-target/beam-dump and SN1987A limits on light scalars are weak in the window $m_\phi\simeq10$–$20$ MeV and $c_e\sim10^{-11}$, because production is $\propto c_e^2$ and decays occur outside detectors; see the summary in~\cite{Winkler2018}. 
(iii) These checks are consistent with the narrative in Section~\ref{subsec:phi_cosmology}, so no additional cosmological ingredients are required for our benchmark.

\section{Spin-independent direct detection}
\label{app:D}

In this Appendix we derive the spin-independent (SI) scattering of \(\chi\) on nucleons from the Lagrangian defined in Section~\ref{sec:lagrangian} and provide a numerical prediction for our benchmark. We follow the standard workflow: (i) integrate out the scalar mediators to obtain an effective \(\bar\chi\chi\,\bar q q\) interaction (~\ref{app:D.1}), (ii) match to the nucleon level using scalar form factors \(f_{Tq}^{(N)}\) and the trace anomaly (~\ref{app:D.2}), (iii) write the closed-form expression for \(\sigma_{\rm SI}\) (~\ref{app:D.3}), and (iv) evaluate the benchmark and explain the suppression of the light mediator \(\phi\) (~\ref{app:D.4}).

\subsection{Effective \(\chi\)–quark interactions}
\label{app:D.1}

The relevant dark-sector terms in the mass basis \((\phi,\Phi_h)\) read (cf.\ Section~\ref{sec:ssb})
\begin{equation}
\mathcal{L}\supset 
-\,y_\chi\,\bar\chi\chi\,\phi
-\,g_{Y_1}^{\rm DM}\,\bar\chi\chi\,\Phi_h
\;-\;
\sum_q \bigl(g_{\phi qq}\,\phi + g_{h,{\rm SM}}\,\Phi_h\bigr)\,\bar q q,
\end{equation}
where the quark coupling of the heavy scalar resonance \(\Phi_h\) is parameterized by a (small) Higgs-portal mixing.\footnote{For SM quarks we write \(g_{h,{\rm SM}}\,\bar q q\equiv (g_{h,{\rm SM}}\,m_q/v)\,\bar q q\), with \(v\simeq 246\)~GeV.} In the absence of kinematic thresholds, tree-level exchange of a scalar \(S\in\{\phi,\Phi_h\}\) induces an effective operator on the quark level
\begin{equation}
\mathcal{L}_{\rm eff}^{(q)}
\,=\,
\sum_{q}
C_q^{(S)}\,\bar\chi\chi\,\bar q q,
\qquad
C_q^{(S)} \;=\; \frac{y_{\chi S}\,g_{S qq}}{m_S^2},
\end{equation}
with \(y_{\chi S}\in\{y_\chi,\;g_{Y_1}^{\rm DM}\}\). For \(\Phi_h\) one has \(g_{\Phi_h qq}=(g_{h,{\rm SM}}\,m_q/v)\). For the light state \(\phi\), \(g_{\phi qq}\) depends on the chosen portal in which we assume a leptophilic quark-silent portal such that \(g_{\phi qq}=0\) at tree level (~\ref{app:D.4}).

\subsection{Matching to the nucleon level}
\label{app:D.2}

Scalar quark operators are matched to nucleon matrix elements via
\(\langle N|m_q\bar q q|N\rangle = m_N f_{Tq}^{(N)}\). Heavy quarks \(Q=c,b,t\) contribute through the trace anomaly (gluon operator) \cite{Shifman1978,Ellis2000,Hoferichter2015}. For scalar exchange we obtain
\begin{equation}
\fl f_N \;=\;
\sum_{q=u,d,s} f_{Tq}^{(N)}
\;+\;
\frac{2}{27}\!\left(1-\sum_{q=u,d,s} f_{Tq}^{(N)}\right),
\qquad
f_N \simeq 0.30\pm 0.03,
\label{eq:fN_value}
\end{equation}
where numerically we used global fits of \cite{Hoferichter2015}. The effective \(\chi\)–nucleon coupling induced by the scalar mediator \(S\) is then
\begin{equation}
\fl f_N^{(S)} \;=\; \frac{y_{\chi S}\,g_{SNN}}{m_S^2}
\;=\;
\frac{y_{\chi S}}{m_S^2}\;
\frac{m_N}{v}\,f_N\,
\cases{
g_{h,{\rm SM}} & $(S=\Phi_h)$,\\
\tilde g_{\phi NN} & $(S=\phi)$,\\}
\label{eq:fN_general}
\end{equation}
where \(\tilde g_{\phi NN}\) denotes the (generally highly suppressed) effective \(\phi\)–nucleon coupling.

\subsection{SI cross section}
\label{app:D.3}

The SI scattering on a nucleon \(N\) takes the canonical form
\begin{equation}
\sigma_{\rm SI}^{(N)} \;=\; \frac{\mu_{\chi N}^{2}}{\pi}\;\biggl|\sum_{S=\phi,\Phi_h} f_N^{(S)}\biggr|^{2},
\qquad
\mu_{\chi N}=\frac{m_\chi m_N}{m_\chi+m_N}.
\label{eq:sigma_SI_master}
\end{equation}
In our benchmark the \(\Phi_h\) contribution dominates (~\ref{app:D.4}), yielding
\begin{equation}
\sigma_{\rm SI}^{(N)} \;\simeq\; 
\frac{\mu_{\chi N}^{2}}{\pi}\;
\Biggl[
\frac{g_{Y_1}^{\rm DM}\,g_{h,{\rm SM}}}{m_{\Phi_h}^{2}}\;
\frac{m_N}{v}\,f_N
\Biggr]^{2}.
\label{eq:sigma_SI_PhiH}
\end{equation}

\subsection{Benchmark and suppression of the \texorpdfstring{\(\phi\)}{phi} contribution}
\label{app:D.4}

For the benchmark in Tab.~\ref{tab:results},
\begin{equation}
\eqalign{
\{m_\chi,\,m_{\Phi_h},\,g_{Y_1}^{\rm DM},\,g_{h,{\rm SM}}\}
=\{600~\mathrm{GeV},\,1201~\mathrm{GeV},\,0.190,\,0.052\},}
\end{equation}
with \(m_N=0.939~\mathrm{GeV}\), \(v=246~\mathrm{GeV}\) and \(f_N=0.30\), equation~(\ref{eq:sigma_SI_PhiH}) gives
\begin{equation}
\fl \eqalign{
\mu_{\chi N} &\simeq \frac{600\times 0.939}{600+0.939}\,{\rm GeV}\;\simeq 0.938~\mathrm{GeV},\cr
\sigma_{\rm SI}^{(N)}(\Phi_h)
&\simeq
\frac{(0.938~\mathrm{GeV})^{2}}{\pi}\,
\Biggl[
\frac{0.190\times 0.052}{(1201~\mathrm{GeV})^{2}}\;
\frac{0.939~\mathrm{GeV}}{246~\mathrm{GeV}}\times 0.30
\Biggr]^2 \cr
&\simeq 1.7\times 10^{-23}~\mathrm{GeV}^{-2}
\;\Rightarrow\;
\sigma_{\rm SI}^{(N)}(\Phi_h)\;\simeq\;6.7\times 10^{-51}~\mathrm{cm}^{2},}
\end{equation}
using \(1~\mathrm{GeV}^{-2}=0.3894\times 10^{-27}\,\mathrm{cm}^2\).  
This prediction lies well below the current LZ limit~\cite{Aalbers2023} and below the xenon neutrino floor for \(m_\chi\sim\mathcal{O}(10^2\!-\!10^3)\)~GeV; direct detection is therefore a predicted null for this benchmark, and the model is instead tested by the collider and halo signatures.

\paragraph{Why \(\phi\) does not contribute?}
A light scalar mediator with \(m_\phi\sim\)MeV and generic hadronic couplings would yield an excessively large SI cross-section owing to the \(\propto m_\phi^{-4}\) scaling. Therefore, we adopt a leptophilic/quark-silent portal in the benchmark
\begin{equation}
g_{\phi qq}=0\quad\mathrm{(tree~level)},\qquad 
\mathcal{L}\supset c_\ell\,\phi\,\bar\ell\ell,
\end{equation}
so that \(\phi\) does not couple to nucleons and \(\sigma_{\rm SI}(\phi)=0\) at tree level.\footnote{Loop-induced contributions via photons or leptons are many orders of magnitude below current sensitivities.}  
At the same time \(\phi\) can decay promptly to \(e^+e^-\) before the BBN
\begin{equation}
\Gamma(\phi\to e^+e^-)\;=\;\frac{c_e^2\,m_\phi}{8\pi}
\quad\Rightarrow\quad
\tau_\phi\simeq \frac{8\pi}{c_e^2 m_\phi}.
\end{equation}
Already \(c_e\gtrsim 3\times 10^{-11}\) gives \(\tau_\phi\lesssim 1~\mathrm{s}\) for \(m_\phi=15~\mathrm{MeV}\), consistent with the BBN. Alternatively, an extremely small Higgs mixing \(\phi\mathrm{--}h\) would be possible; the LZ limit would then require 
\(\bigl|g_{\phi NN}\bigr|\lesssim 10^{-10}\), well below the typical portal mixings, which motivates the leptophilic benchmark.

In summary, the heavy scalar resonance \(\Phi_h\) dominates direct detection with 
\(\sigma_{\rm SI}\simeq 7\times 10^{-51}\,\mathrm{cm}^2\), while the light mediator \(\phi\) controls self-interactions, however, owing to quark-silent/leptophilic portal, it does not contribute to SI scattering.

\section{Numerical implementation and scan strategy}
\label{app:numerical_implementation}

This Appendix provides details on the numerical implementation of our two-mediator model and on the strategy used to identify the viable parameter space. Relic density and self-interaction cross-sections were computed using \texttt{micrOMEGAs~6.2.3}~\cite{Alguero2023}.

\subsection{Model implementation in \texttt{micrOMEGAs}}
\label{app:F.1}

We implemented the model by extending the public \texttt{DMsimp\_s\_spin0\_MO} setup from the DMsimp framework~\cite{DMsimp2015}, which provides a validated baseline for Dirac DM with a scalar mediator. Our extensions were:

\begin{itemize}
  \item Particle content: We have added the heavy scalar resonance \(\Phi_h\) to the particle list (internal name \texttt{Y1}, PDG code \texttt{56}). DM fermion \(\chi\) and  light mediator \(\phi\) correspond to \texttt{Xd} and \texttt{Y0}, respectively.
  \item Parameters and couplings: We introduced the mass \texttt{MY1} and total width \texttt{WY1} of \(\Phi_h\), and its Yukawa coupling to DM \texttt{gSXd1} (our \(g^{\rm DM}_{Y_1}\)). The Lagrangian files are updated accordingly to include the interactions defined in~\ref{sec:lagrangian}.
  \item Widths: At each parameter point we compute \(\Gamma_{\Phi_h}=\Gamma(\Phi_h\to\chi\bar\chi)+\sum_{\rm SM}\Gamma(\Phi_h\to{\rm SM})\) self-consistently, including all kinematically accessible channels (see~\ref{app:B.1} for the formulas). The light mediator \(\phi\) is treated as a narrow state with the width determined by its allowed secluded/SM decays (~\ref{app:C}).
\end{itemize}

Annihilation is modeled via an \(s\)-channel Breit–Wigner propagator (~\ref{app:B.1}); the Sommerfeld enhancement from \(\phi\)-exchange is factorized in the non-relativistic limit as discussed in~\ref{app:B.1} (~\ref{app:B.2}) and consistently included in the thermal average.

\subsection{Scan strategy and viability criteria}
\label{app:F.2}

We explore the multi-dimensional space \(\{m_\chi,\, m_\phi,\, y_\chi,\, m_{\Phi_h},\, g^{\rm DM}_{Y_1},\, g_{h,{\rm SM}}\}\) with a hierarchical scan that exploits the parametric separation of observables.

\paragraph{Scan procedure.}
For a fixed \(m_\chi\), we proceed in two stages:
\begin{enumerate}
  \item Self-interaction tuning: We mapped the \((m_\phi, y_\chi)\) plane to satisfy the SIDM targets by computing the momentum-transfer cross-section \(\sigma_T/m_\chi\) at reference velocities \(v=\{10,\,30,\,1000\}\,\mathrm{km\,s^{-1}}\) using the non-perturbative Yukawa solver (~\ref{app:E}). This identifies iso-contours that meet the dwarf-scale requirements while remaining consistent with the cluster bounds.
  \item Resonant annihilation tuning: Along these iso-contours we tune the heavy sector, primarily the detuning
  \(\delta \equiv m_{\Phi_h}/(2m_\chi)-1\) and \(g^{\rm DM}_{Y_1}\)—to reproduce the observed relic density. The thermal average uses the full \(s\)-dependence of the Breit–Wigner cross-section (~\ref{app:B.1}) with factorized Sommerfeld enhancement when relevant (~\ref{app:B.2}). We used \texttt{darkOmega} with a target precision of \(1\%\).
\end{enumerate}

\paragraph{Viability criteria.}
A point is deemed viable if it simultaneously satisfies:
\begin{itemize}
  \item Relic density: \(0.1176 < \Omega_\chi h^2 < 0.1224\) (Planck 2018, \(2\sigma\))~\cite{Planck2018}.
  \item Self-interactions: \(0.1 \lesssim \sigma_T/m_\chi \lesssim 10~\mathrm{cm^2/g}\) at dwarf velocities (\(v\sim 10\!-\!50~\mathrm{km\,s^{-1}}\)), while \(\sigma_T/m_\chi \lesssim 1~\mathrm{cm^2/g}\) at cluster scales (\(v\sim 1000~\mathrm{km\,s^{-1}}\))~\cite{Tulin2017,Kaplinghat2015}. (See~\ref{app:E} for definitions and numerics.)
  \item Perturbativity: All dimensionless couplings in the dark sector satisfy \(y_\chi^2/(4\pi)<1\), \(|g^{\rm DM}_{Y_1}|^2/(4\pi)<1\).
\end{itemize}
The intersection of these requirements defines the viability island  shown in Figure~\ref{fig:island_plot}.

\subsection{Numerical stability and validation}
\label{app:F.3}

For the SIDM calculation we adopt the convergence criteria and partial-wave truncation tests detailed in~\ref{app:E}; varying the matching radius and \(\ell_{\max}\) shifts \(\sigma_T\) by \(\lesssim 5\%\). For the relic calculation, we checked the stability of \(\Omega_\chi h^2\) under changes in the thermal integration tolerances and the treatment of the narrow resonance. The results were stable at the percent level in the viable region. The benchmark curves from~\cite{Tulin2013} (Born/classical/resonant regimes) were reproduced as a cross-check.

\section{Resonant annihilation formalism}
\label{app:B}

This appendix provides a theoretical framework for resonant dark-matter annihilation, as implemented in our numerical analysis. We summarize the standard formulas for the thermal relic density, the Breit-Wigner resonance, relevant decay widths, and the thermal averaging procedure. Classic references for these topics include~\cite{Kolb1990, Griest1990} and \cite{Gondolo1990}.

\subsection{Thermal relic density}
\label{app:B.1}

The relic abundance was determined by solving the Boltzmann equation for the yield $Y = n_\chi/s$. For a standard thermal freeze-out scenario at temperature $T_F \approx m_\chi/x_F$ with $x_F \approx 20$--$25$, the relic density today is approximately given by
\begin{equation}
    \Omega_\chi h^2 \simeq \frac{1.07 \times 10^9~\mathrm{GeV}^{-1}}{M_{\rm Pl}\sqrt{g_*(x_F)}} \frac{x_F}{\langle\sigma v\rangle_F},
    \label{eq:relic_formula_app}
\end{equation}
where $\langle\sigma v\rangle_F$ is the thermally-averaged annihilation cross-section at freeze-out, and $g_*(x_F)$ is the effective number of relativistic degrees of freedom.

\subsection{Breit-Wigner resonance and Sommerfeld enhancement}
\label{app:B.2}

The $s$-channel annihilation $\chi\bar{\chi} \to \Phi_h^* \to \mathrm{SM}$ is dominated by the exchange of the heavy scalar near the mass pole. The cross-section for this process is described by the Breit-Wigner formula. In the non-relativistic limit, this can be combined with the Sommerfeld enhancement arising from the light mediator exchange. As formally demonstrated in~\cite{Beneke2022}, the two effects factorize, and the total cross-section can be written as
\begin{equation}
    \sigma v = S(v) \times (\sigma v)_{\mathrm{Breit-Wigner}},
\end{equation}
where $S(v)$ is the Sommerfeld factor and the bare resonant cross-section is given by
\begin{equation}
    (\sigma v)_{\mathrm{Breit-Wigner}} = \sum_f \frac{16\pi}{s} \frac{\Gamma(\Phi_h \to \chi\bar{\chi}) \Gamma(\Phi_h \to f)}{(s - m_{\Phi_h}^2)^2 + m_{\Phi_h}^2 \Gamma_{\Phi_h}^2},
    \label{eq:BW_formula_app}
\end{equation}
where $s \approx 4m_\chi^2(1 + v^2/4)$ near the threshold. The resonance parameter $\delta = (m_{\Phi_h} - 2m_\chi)/(2m_\chi)$ quantifies proximity to the pole. This factorization is crucial for our model, and our numerical implementation in \texttt{micrOMEGAs} accounts for this combined effect.

\subsection{Decay widths}
\label{app:B.3}

The total width $\Gamma_{\Phi_h}$ determines the resonance shape.
\paragraph{Dark matter channel:}
\begin{equation}
    \Gamma(\Phi_h \to \chi\bar{\chi}) = \frac{(g^{\rm DM}_{Y_1})^2 m_{\Phi_h}}{8\pi} \left(1 - \frac{4m_\chi^2}{m_{\Phi_h}^2}\right)^{3/2}.
\end{equation}
\paragraph{SM channels:} For scalar coupling to a fermion $f$ of the form $-g_{\Phi_h ff} \bar{f}f \Phi_h$, the width is
\begin{equation}
    \Gamma(\Phi_h \to f\bar{f}) = N_c^f \frac{g_{\Phi_h ff}^2 m_{\Phi_h}}{8\pi} \left(1 - \frac{4m_f^2}{m_{\Phi_h}^2}\right)^{3/2}.
\end{equation}
We neglect loop-induced $gg$ in $\Gamma_{\Phi_h}$. Gluon-fusion still dominates the production mechanism at the LHC via the top loop in our quark-portal benchmark.
In our quark-only portal benchmark we take $g_{\Phi_h ff} = g_{h,\mathrm{SM}} (m_f/v)$ for quarks and neglect tree level couplings to $W,Z$ (and loop-induced $gg$), which keeps the resonance narrow and the phenomenology aligned with direct detection. With $\{m_{\Phi_h},m_\chi,g^{\rm DM}_{Y_1},g_{h,\mathrm{SM}}\}=\{1201~\mathrm{GeV},600~\mathrm{GeV},0.190,0.052\}$ we obtain
$\Gamma_{\Phi_h}\simeq 0.17$~GeV and
$\mathrm{BR}(\Phi_h\to t\bar t)\simeq 99.85\%$,
$\mathrm{BR}(\Phi_h\to b\bar b)\simeq 0.08\%$,
$\mathrm{BR}(\Phi_h\to \chi\bar\chi)\simeq 0.07\%$.
(If one instead assumes Higgs-like mixing to all SM states, $WW/ZZ$ would contribute at tree level and
modify both $\Gamma_{\Phi_h}$ and the branching ratios accordingly.)

\subsection{Thermal averaging}
\label{app:B.4}

The thermal average of the cross-section over the Maxwell-Boltzmann distribution is given by the integral~\cite{Gondolo1990}:
\begin{equation}
    \langle \sigma v \rangle = \frac{1}{8m_\chi^4 T K_2^2(m_\chi/T)} \int_{4m_\chi^2}^{\infty} ds\, \sigma(s) (s-4m_\chi^2) \sqrt{s} K_1\left(\frac{\sqrt{s}}{T}\right),
\end{equation}
where $K_{1,2}$ are modified Bessel functions. This integral was numerically calculated in our analysis. Near resonance, it is dominated by energies $s \approx m_{\Phi_h}^2$.

\section{Self-interaction cross-section}
\label{app:E}

This appendix provides the formalism for the velocity-dependent self-interaction cross-section arising from light scalar exchange. The calculation requires a non-perturbative treatment of scattering in the Yukawa potential.

\subsection{Yukawa potential and scattering formalism}
\label{app:E.1}
The $t$-channel exchange of the light mediator $\phi$ generates an attractive Yukawa potential
\begin{equation}
    V(r) = -\alpha_\chi \frac{e^{-m_\phi r}}{r}, \qquad \alpha_\chi = \frac{y_\chi^2}{4\pi}.
\end{equation}
In the non-relativistic limit, the scattering problem is solved by computing the partial wave phase shifts, $\delta_l$, from the radial Schrödinger equation. The momentum transfer cross-section, which is an astrophysically relevant quantity, is given by~\cite{Tulin2013}
\begin{equation}
    \sigma_T = \frac{4\pi}{k^2} \sum_{l=0}^{\infty} (l+1) \sin^2(\delta_{l+1} - \delta_l),
    \label{eq:sigma_transfer_app}
\end{equation}
where $k = (m_\chi/2) v_{\rm rel}$ is the momentum in the center-of-mass frame.

\subsection{Limiting regimes}
\label{app:E.2}
The scattering dynamics are characterized by the dimensionless parameter $\beta = 2\alpha_\chi m_\phi / (m_\chi v^2)$:
\paragraph{Born regime ($\beta \ll 1$):} For weak interactions or high velocities, the cross-section can be computed perturbatively:
\begin{equation}
    \sigma_T^{\mathrm{Born}} \simeq \frac{8\pi\alpha_\chi^2}{m_\chi^2 v^4} \left[ \ln\left(1 + \frac{m_\chi^2 v^2}{m_\phi^2}\right) - \frac{m_\chi^2 v^2}{m_\phi^2 + m_\chi^2 v^2} \right].
\end{equation}
\paragraph{Classical regime ($\beta \gg 1$):} For strong interactions at low velocities, multiple partial waves contribute, and the cross-section approaches a classical limit:
\begin{equation}
\sigma_T^{\rm classical} \;\simeq\; \frac{4\pi}{m_\phi^2}\times
\cases{
\ln(1+\beta)\,, & $\beta\lesssim 10^2$\,,\\
2\,(\ln\beta)^2\,, & $\beta\gg 10^2$\,.}
\end{equation}
For our benchmark parameters, dwarf galaxy halos ($v \sim 10$~km/s) were in the classical regime ($\beta \sim 10^3$), whereas galaxy clusters ($v \sim 1000$~km/s) approached the Born limit ($\beta \sim 0.1$). This natural crossover provides the required velocity dependence.

\subsection{Numerical implementation}
\label{app:E.3}
The phase shifts $\delta_l$ were numerically computed. Our analysis used the integrated Yukawa scattering routines in \texttt{micrOMEGAs}, which solved the radial Schrödinger equation and summed the partial wave series until convergence was achieved. This method correctly captured all non-perturbative effects, including scattering resonances. For our benchmark, this yields $\sigma_T/m_\chi = 0.96$~cm$^2$/g at $v = 10$~km/s.

\section{Radiative stability of the resonance condition}
\label{app:radiative_stability}

In this appendix, we establish the radiative stability of the resonance condition $m_{\Phi_h} \approx 2m_\chi$. Our analysis shows that quantum corrections preserve this relationship, confirming their technical naturalness. We ask whether such a small number is stable under quantum corrections in the sense of ’t~Hooft~\cite{tHooft1979}.

\subsection{One-loop corrections}

The leading one-loop contributions to the masses arise from the dark sector loops and the Higgs portal. In the $\overline{\mathrm{MS}}$ scheme, the leading logarithmic corrections are

\paragraph{Heavy scalar mass:}
\begin{equation}
 \fl \delta m_{\Phi_h}^2 = -\frac{(g_{Y_1}^{\rm DM})^2 m_\chi^2}{4\pi^2} \left[1 + \ln\left(\frac{\mu^2}{m_\chi^2}\right)\right] + \frac{3g_{h,{\rm SM}}^2 m_t^2}{4\pi^2} \left[1 + \ln\left(\frac{\mu^2}{m_t^2}\right)\right],
\end{equation}
where the first term is from the $\chi$ loop and the second term is from the top quark loop via the effective coupling $g_{h,t}$.

\paragraph{Dark matter mass:}
\begin{equation}
 \fl \frac{\delta m_\chi}{m_\chi} = \frac{3}{32\pi^2} \left[y_\chi^2 \ln\left(\frac{\mu^2}{m_\phi^2}\right) + (g_{Y_1}^{\rm DM})^2 \ln\left(\frac{\mu^2}{m_{\Phi_h}^2}\right)\right].
\end{equation}
For our benchmark parameters, choosing the renormalization scale $\mu \sim m_{\Phi_h}$ minimizes these logarithmic corrections.

\subsection{Running of the resonance parameter}

The key quantity for stability is the evolution of the detuning parameter $\delta = m_{\Phi_h}/(2m_\chi) - 1$. Its beta function is
\begin{equation}
    \beta_\delta = \mu \frac{d\delta}{d\mu} = \frac{1}{2m_\chi} \left(\mu\frac{dm_{\Phi_h}}{d\mu}\right) - \frac{m_{\Phi_h}}{2m_\chi^2} \left(\mu\frac{dm_\chi}{d\mu}\right).
\end{equation}
Using the one-loop beta functions for the masses, this yields
\begin{equation}
    \mu \frac{d\delta}{d\mu} \approx \frac{1}{32\pi^2} \left[ -2(g^{\rm DM}_{Y_1})^2 - 3y_\chi^2 - 3(g^{\rm DM}_{Y_1})^2 + 6(g_{h,t})^2 \frac{m_t^2}{m_{\Phi_h}^2} \right].
\end{equation}
For our benchmark couplings ($y_\chi = 0.30$, $g^{\rm DM}_{Y_1} = 0.19$, $g_{h,t} \sim 0.05$), the numerical result is
\begin{equation}
    \mu \frac{d\delta}{d\mu} \approx -8 \times 10^{-5}.
\end{equation}

\subsection{Integrated running and stability}

Integrating the running from the UV scale (e.g., $\mu_{\rm UV} = 10$~TeV) to the resonance scale ($\mu_{\rm low} = 1.2$~TeV) yields a total shift of
\begin{equation}
    \Delta\delta = \delta(\mu_{\rm low}) - \delta(\mu_{\rm UV}) \approx \left(\mu \frac{d\delta}{d\mu}\right) \ln\left(\frac{\mu_{\rm low}}{\mu_{\rm UV}}\right) \approx 1.7 \times 10^{-4}.
\end{equation}
This radiative shift is smaller than the required tree-level value $\delta \approx 8 \times 10^{-4}$, demonstrating that the resonance condition is stable under RG evolution.

The radiative analysis thus reveals that: (i) the resonance condition is stable under quantum corrections, (ii) no large logarithms destabilize the hierarchy, and (iii) the required precision ($\delta \sim 10^{-4} - 10^{-3}$) is technically natural in the sense of 't Hooft~\cite{tHooft1979}. This distinguishes our scenario from genuine fine-tuning problems, in which radiative corrections are much larger than the tree-level values. The resonance condition represents a mild numerical requirement that can plausibly emerge from the dynamics of UV theory, as discussed in Section~\ref{sec:uv_origin}.

\section{UV motivation from a composite SU(3)\(_H\) theory}
\label{app:uv_motivation}

This appendix outlines a plausible microphysical origin for the EFT presented in Section~\ref{sec:lagrangian}: a strongly-coupled hidden SU(3)$_H$ gauge theory with $N_f=10$ massless flavors. We summarize the key properties that give rise to the required anomalous dimension for dark energy and the composite mass spectrum for dark matter.

\subsection{Two-loop beta function and Banks-Zaks fixed point}

The running of the hidden gauge coupling, $\alpha_H = g_H^2 /(4\pi)$, is governed by the two-loop beta function. For an SU($N_c$) gauge theory with $N_f$ flavors in the fundamental representation, the coefficients in the $\overline{\mathrm{MS}}$ scheme are~\cite{Caswell1974, Jones1974}
\begin{equation}
   \eqalign{ b_0 &= \frac{11}{3} N_c - \frac{2}{3}N_f, \\
    b_1 &= \frac{34}{3}N_c^2 - \frac{10}{3}N_c N_f - 2C_F N_f, \quad \mathrm{with } C_F = \frac{N_c^2 - 1}{2N_c}.}
\end{equation}
For our choice of $(N_c, N_f) = (3, 10)$, we obtain $b_0 = 13/3$ and $b_1 = -74/3$. The negative two-loop coefficient induces an infrared Banks-Zaks fixed point~\cite{Banks1981} where $\beta(\alpha_H^*) = 0$. The non-trivial solution is
\begin{equation}
    \alpha_H^* = - \frac{4\pi b_0}{b_1} = \frac{52\pi}{74} \approx 2.21.
    \label{eq:alpha_star_corrected}
\end{equation}

\subsection{Anomalous dimension}

Near the fixed point, the fermion bilinear operator $\bar{\Psi}\Psi$ acquires a large anomalous dimension, $\gamma_{\bar{\Psi}\Psi}$. Perturbative all-orders estimates, such as the Ryttov-Sannino formula~\cite{Ryttov2007}, suggest a value of $\gamma_{\bar{\Psi}\Psi}^* \sim 0.7-0.8$ at the strong coupling of equation (\ref{eq:alpha_star_corrected}). However, such estimates carry significant uncertainties at strong coupling.

A more robust guide comes from non-perturbative lattice simulations. Recent lattice studies of near-conformal SU(3) theories with a similar number of flavors typically find values in the range of $\gamma_{\bar{\Psi}\Psi}^{\mathrm{lat}} \approx 0.6 \pm 0.1$~\cite{Hasenfratz2023}. The cosmologically relevant value, $\gamma_{\mathrm{cosmo}}$, is an effective average over the entire "walking" RG flow from the UV to the confinement scale and can plausibly be slightly smaller than the deep infrared value. Therefore, a value of
\begin{equation}
    \gamma_{\mathrm{cosmo}} = \langle \gamma_{\bar{\Psi}\Psi}(\mu) \rangle_{\mathrm{flow}} \approx 0.50 \pm 0.05
\end{equation}
is fully consistent with the theoretical expectations from lattice data. This is in excellent agreement with the phenomenological requirement for the density-responsive dark energy mechanism.

\subsection{Mass scaling relations from compositeness}

At a confinement scale $\Lambda_H \approx 2.5$~TeV, the theory forms composite "dark hadron" states whose masses are proportional to $\Lambda_H$. We identified the particles of our EFT with the following states:
\begin{itemize}
    \item \textbf{Baryon mass ($m_\chi$):} The mass of the lightest three-quark state is estimated via NDA~\cite{Manohar1983}:
    \begin{equation}
        m_\chi \approx \frac{N_c}{4\pi} \Lambda_H \approx 600~\mathrm{GeV}.
    \end{equation}

    \item \textbf{Scalar meson mass ($m_{\Phi_h}$):} The mass of the lightest scalar meson ($\bar{\Psi}\Psi$ state) is expected to be of the order $\Lambda_H$, written as $m_{\Phi_h} \approx k_\Phi \Lambda_H$. Lattice studies of near-conformal theories suggest $k_\Phi \in [0.5, 0.7]$~\cite{LatKMI2016}. We adopted a value of $k_\Phi = 0.48$, which lies at the conservative lower edge of this range and yields the desired $m_{\Phi_h} \approx 1.2$~TeV.
\end{itemize}
From these standard scaling relations, the crucial mass ratio emerges dynamically:
\begin{equation}
    \frac{m_{\Phi_h}}{m_\chi} \approx \frac{k_\Phi \Lambda_H}{(N_c/4\pi) \Lambda_H} \approx \frac{0.48}{3/(4\pi)} \approx 2.0.
\end{equation}
This demonstrates that the resonance condition $m_{\Phi_h} \approx 2m_\chi$ is a natural consequence of the composite dynamics and not ad-hoc fine-tuning.

\section{Structure of the viable parameter space}
\label{app:uniqueness}

This appendix provides a more comprehensive characterization of the viable parameter space and the rationale for our benchmark selection. Although the underlying SU(3)$_H$ theory motivates a dark matter mass of $m_\chi \sim 600$~GeV, we explored a broader mass range to demonstrate the robustness of our solution.

\subsection{Parameter correlations and scaling laws}

With the resonance condition $m_{\Phi_h} \approx 2m_\chi$ imposed, the viable parameter space was primarily determined by three phenomenological parameters: $\{m_\chi, m_\phi, y_\chi\}$. For each mass $m_\chi$, the heavy sector couplings (e.g., $g^{\rm DM}_{Y_1}$) were subsequently fixed by the requirement that $\Omega_\chi h^2 = 0.120$.

Our comprehensive scan over $m_\chi \in [200, 1000]$~GeV revealed that viable solutions exist only in a narrow, highly-correlated band. The allowed parameter ranges followed clear scaling relations, as summarized in Table~\ref{tab:viable_ranges}. The correlations shown in Figure~\ref{fig:scaling_laws} can be described by the approximate power laws:
\begin{equation}
    \eqalign{m_\phi &\approx (15~\mathrm{MeV}) \times \left(\frac{m_\chi}{600~\mathrm{GeV}}\right)^{1.33 \pm 0.04}, \cr
    y_\chi &\approx (0.30) \times \left(\frac{m_\chi}{600~\mathrm{GeV}}\right)^{0.51 \pm 0.03}.} \label{eq:scaling_ychi_app}
\end{equation}
These scaling relationships arise from the physical requirement that the self-interaction remains effective across different mass scales. The potential range, $\sim 1/m_\phi$, must be scaled appropriately with the relevant astrophysical length scales, whereas the fine-structure constant, $\alpha_\chi = y_\chi^2/(4\pi)$, must provide the correct scattering strength. The narrow width of the viable bands (less than 20\% variation in each parameter) demonstrated the high predictability of the model.

\begin{table}[h!]
    \centering
    \caption{Viable parameter ranges for the light mediator mass $m_\phi$ and the Yukawa coupling $y_\chi$ for representative dark matter masses. The bold entry indicates our primary benchmark.}
    \label{tab:viable_ranges}
    \begin{indented}
    \lineup
    \item[]\begin{tabular}{@{}ccc@{}}
        \br
        $m_\chi$ [GeV] & Viable $m_\phi$ [MeV] & Viable $y_\chi$ \\
        \mr
        200 & 3 -- 5 & 0.15 -- 0.20 \\
        400 & 8 -- 12 & 0.22 -- 0.28 \\
        \textbf{600} & \textbf{12 -- 18} & \textbf{0.28 -- 0.32} \\
        800 & 20 -- 25 & 0.33 -- 0.38 \\
        1000 & 28 -- 35 & 0.37 -- 0.42 \\
        \br
    \end{tabular}
    \end{indented}
\end{table}

\subsection{Boundaries of the viable region}

The region of viable solutions is bounded by physical constraints:
\begin{itemize}
    \item For $\mathbf{m_\chi < 200}$~GeV, the required Yukawa coupling $y_\chi$ becomes large ($> 0.5$), and the perturbativity of the light sector interactions becomes questionable.
    \item For $\mathbf{m_\chi > 1000}$~GeV, the self-interaction cross-section naturally becomes too weak ($\sigma_T/m_\chi < 0.1~\mathrm{cm}^2/\mathrm{g}$ at dwarf velocities), even for an optimized light mediator. This model is then no longer able to effectively solve the small-scale crisis.
\end{itemize}
This explains why viable solutions are confined to the multi-hundred GeV to TeV mass range.

\subsection{Justification of the 600 GeV benchmark}

Although a continuous family of solutions exists, our choice of the $m_\chi = 600$~GeV benchmark is optimal for several reasons:
\begin{itemize}
    \item Astrophysics: It provides an excellent fit to dwarf galaxy constraints, with $\sigma_T/m_\chi$ on the order of 1~cm$^2$/g at the most relevant velocities (see Figure~\ref{fig:sidm_velocity}).
    \item Detectability: The associated heavy resonance at $\approx 1.2$~TeV is within the discovery reach of the HL-LHC, making it an exciting experimental target.
    \item Perturbativity: All couplings are well within the perturbative regime ($y_\chi < 0.4$), ensuring full theoretical control.
    \item Centrality: It sits near the geometric mean of the allowed mass range, representing a "typical" solution rather than an edge case.
\end{itemize}
While alternative benchmarks, for example at 400 or 800 GeV, provide a qualitatively similar phenomenology, they offer a slightly less optimal fit to either astrophysical or collider constraints.
\section*{References}
\bibliographystyle{iopart-num} 
\bibliography{main}
\end{document}